\documentclass[]{IEEEtran}
\usepackage{cite}
\usepackage{amsmath,amssymb,amsfonts,mathtools}
\usepackage{bbm, bm}
\usepackage{algorithm}
\usepackage{algorithmic}
\usepackage{graphicx}
\usepackage{textcomp}
\usepackage{xcolor}
\usepackage{subcaption}
\usepackage{siunitx}
\usepackage{hyperref}
\usepackage{flushend}

\newcommand{\grad}{\nabla}

\newcommand{\bI}{\boldsymbol{I}}

\newcommand{\bzero}{\boldsymbol{0}}

\newcommand{\bx}{\boldsymbol{x}}

\newcommand{\bz}{\boldsymbol{z}}

\newtheorem{remark}{Remark}

\def\BibTeX{{\rm B\kern-.05em{\sc i\kern-.025em b}\kern-.08em
    T\kern-.1667em\lower.7ex\hbox{E}\kern-.125emX}}
    
\definecolor{mehdi}{RGB}{0,0,250}

\definecolor{Samad}{RGB}{0,250,0}

\definecolor{Matti}{RGB}{250,0,0}

\graphicspath{ {./Figures/} } 
    
\begin{document}


\title{Conditional Denoising Diffusion Probabilistic  Models for Data Reconstruction  Enhancement in Wireless Communications}

\markboth{}
{}
\author{\IEEEauthorblockN
{Mehdi Letafati, \IEEEmembership{Student Member, IEEE,}
			Samad Ali, \IEEEmembership{Member, IEEE,}   and
		Matti Latva-aho,
		   	\IEEEmembership{Fellow, IEEE}
}
\textsuperscript{}\thanks{
Preliminary results of this paper were  presented at the IEEE Wireless Communications and Networking Conference (WCNC 2024), Dubai, United Arab Emirates, Apr. 2024 \cite{conf_hwi}.  

The authors are with the Centre for Wireless Communications, University of Oulu,
Oulu, Finland (e-mails: mehdi.letafati@oulu.fi; 
 samad.ali@oulu.fi; matti.latva-aho@oulu.fi). 
}}


\maketitle

\begin{abstract}
In this paper, 
conditional denoising diffusion probabilistic models (CDiffs)   
are proposed to 
enhance the data  transmission and reconstruction over wireless channels.        
The underlying mechanism of diffusion models  is to decompose the data  generation process over the so-called  ``denoising'' steps.  
Inspired by this, the key idea is to leverage the generative prior of diffusion models  
in learning a ``noisy-to-clean''  transformation of the information signal to help  enhance data reconstruction.  The proposed scheme could be beneficial for communication scenarios in which a prior knowledge of the information content is available, e.g., in multimedia transmission.  Hence, instead of employing complicated channel codes that reduce the information rate, one can exploit diffusion priors for reliable  data reconstruction, especially under extreme channel conditions due to low signal-to-noise ratio (SNR), or hardware-impaired communications. 
 The proposed CDiff-assisted  receiver is 
tailored for the scenario of wireless  image transmission using  MNIST dataset. 
Our numerical results highlight 
the reconstruction  performance of our scheme compared to the conventional digital communication, as well as the deep neural network (DNN)-based benchmark. It is also shown that more than 10 dB improvement in the reconstruction could be achieved in low SNR regimes, 
without the need to reduce  the information rate for error correction. 
\end{abstract}


 \begin{figure*}[]
	\vspace{0mm}
	\centering
	\includegraphics
 [width=
  1.0\textwidth
  ]
 {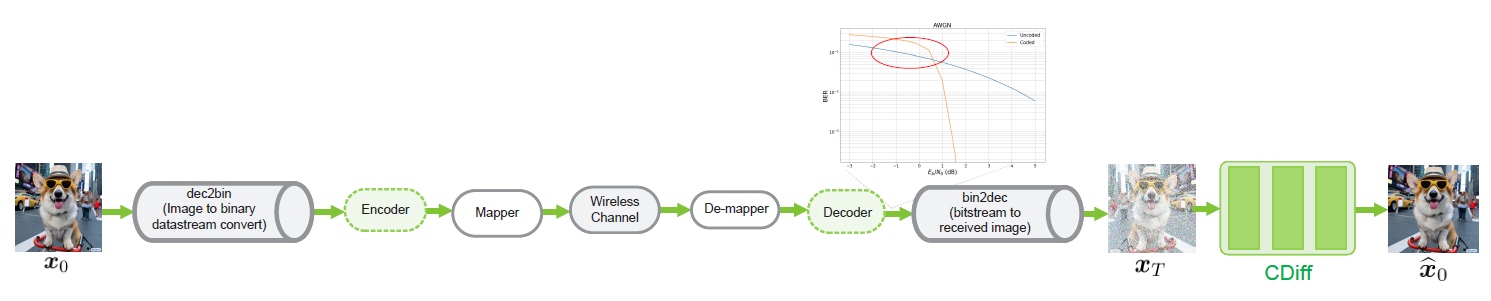}\vspace{-6mm}\caption{\small Data pipeline of our experimental communication  setup for CDiff-based image reconstruction.} 
	\label{fig:pipeline}
 \vspace{0mm}
\end{figure*}

\section{Introduction}  
\subsection{Motivation \& Background} 
With the emergence of deep generative models, the realm of artificial intelligence (AI) has witnessed a paradigm shift  
towards the novel concept of generative AI  that can  facilitate the development of AI-based systems \cite{GenAI_PHY, WiGenAI, wcl}.    
At the same time, from data communication and networking perspective, ``connected intelligence'' is envisioned  as the most significant driving force in the sixth generation (6G) of communications---AI and machine learning (AI/ML) algorithms are envisioned to be widely incorporated into 6G wireless networks, realizing ``AI-native'' communication systems \cite{twelve_6G, 3gpp, hexaX, Samad}.   
This underscores the need for novel AI/ML-based solutions to be tailored to meet the requirements of  emerging communication scenarios.     

The evolution of diffusion models \cite{DM_Ho}, as the new paradigm in deep generative models, can be regarded  as one of the main factors contributing to  the recent breakthroughs in generative AI, that 
have  already showcased notable success with  some of the popular examples 
such as DALL.E 2 by OpenAI,\footnote{https://openai.com/dall-e-2} and
ImageGen by Google Brain.\footnote{https://imagen.research.google/}  Diffusion models have realized unprecedented   results  in different  applications, such as computer vision, natural language processing (NLP), data analysis and synthesis,  and medical imaging \cite{DM_MRI}.  Please see \cite{DM_survey} and references therein for a comprehensive   survey on diffusion models.  
The core concept of diffusion modeling is that {if we could develop an ML model that is capable of learning the \emph{systematic decay of information}, due to noise/distortions,  
then it should be possible to ``reverse'' the process and recover the information back from the noisy/erroneous data.}  
This is fundamentally different from  conventional  generative models, like \textcolor{black}{generative adversarial networks (GANs)} and \textcolor{black}{variational autoencoders (VAEs)}. 
Accordingly,   
{the close underlying relation between the key concepts on how diffusion models work and the   problems in wireless
communication systems has motivated us to carry out this research.} 

\subsection{Related Works}   
Ongoing research  on diffusion models  encompasses both theoretical advancements and practical applications  across  different domains of computer science \cite{DM_survey}. However, there have been only a few papers in wireless communication  literature that have started looking into the potential merits  of  diffusion-based generative  models for wireless systems \cite{DGM_Mag,  DM_for_E2EComm, wcl, WiGenAI, CGM_ChanEst,CGM_ChanEst_WCNC, hybrid, INN}.   
The authors in \cite{DGM_Mag} study a workflow for utilizing  diffusion models in wireless network management.  They    study the incentive mechanisms for generating  contracts using diffusion models in mobile AI-generated content services.    They exploit the 
exploration capability of diffusion models in  Q-learning algorithms for network management.   
Denoising diffusion model is utilized in \cite{DM_for_E2EComm} to generate synthetic channel realizations conditioned on the message information. 
The authors tackle the  problem of 
differentiable channel model within the 
 gradient-based 
training  process of  end-to-end ML-based communications. 
The results highlight the performance  of diffusion models as an alternative to \textcolor{black}{GAN}-based schemes. It is shown in \cite{DM_for_E2EComm} that  \textcolor{black}{GANs}  experience unstable training and less diversity in generation performance,  due to  their adversarial training nature. However,  diffusion models maintain a more stable training process  and a better generalization during inference.  
Noise-conditioned score network (NCSN)-based  channel estimation is studied in \cite{CGM_ChanEst} and \cite{CGM_ChanEst_WCNC}  for  multi-input-multi-output (MIMO) wireless communication scenario. The authors   estimate the gradient of the high-dimensional log-prior of  wireless channels, using RefineNet neural  network architecture.  
Posterior sampling  method is further applied to make the generated  channel estimations consistent with the  pilot signals measured at the receiver. 
Diffusion models have also been incorporated into the so-called semantic communication systems in \cite{INN, hybrid}. 
More specifically, deep learning-based  joint source-channel coding (Deep-JSCC) is combined with diffusion models to  complement semantic data transmission schemes with a generative component.  
The initial results show that the perceptual quality of reconstruction can be improved in such scenarios. 
Nevertheless, those schemes in \cite{INN, hybrid}  propose to employ two different types of deep learning models each  maintaining  a  distinct objective function (while their optimization also  depends on each other), i.e., the autoencoder for semantic transmission and reception,  and the diffusion models for reconstruction enhancement. This  can impose serious concerns in terms of the computational overhead  to the network. In addition,  the output signal of  a neural encoder (i.e., the channel input)  does not necessarily follow the standard format of modulation signals, which makes it challenging to employ semantic transmission ideas in a real-world communication scenario.

\subsection{Our Contributions}  
In this paper, we propose  conditional denoising diffusion probabilistic models (DDPM)   
  for enhancing the wireless  image transmission in  digital communication schemes.   
Different from \textcolor{black}{GANs and VAEs}, the underlying mechanism of diffusion models is to first transform the ground-truth  signal  into an isotropic Gaussian distribution in the forward  diffusion process, and then try to regenerate data  over the so-called  ``denoising'' steps by  
removing the noise introduced in the diffusion process. 
 DDPMs in their vanilla format are  based upon   the assumption that the denoising process starts with  isotropic Gaussian noise \cite{DM_Ho}.     
 However, in a practical wireless system,  the received/decoded signals 
are a degraded version of the ground-truth information signals, not necessarily a pure isotropic Gaussian noise, while the degradation is different in different distortion conditions such as different SNRs and transceiver specifications.    
To address this, we propose  conditional diffusion models (CDiff), in which  the  noisy degraded signals are incorporated  into the denoising framework of the diffusion model. This way,  the diffusion model learns to reconstruct the ground-truth information from the degraded received signal, instead of a pure noise.    

In our scheme, the generative prior of conditioned DDPMs 
is exploited to enhance data reconstruction at the receiver of a  communication system. This is particularly pronounced in extreme channel conditions with poor connectivity  due to low signal-to-noise ratio (SNR), or hardware-impaired communications. In such scenarios, error correcting  codes might not be able to correct the mismatches, or complicated codes might be needed  at the cost of decreasing  the information rate to correct the errors.  
Instead, the proposed diffusion-aided scheme can be utilized  for reliable  data reconstruction without the need for decreasing the information rate.       
This can be tailored for the communication scenarios,  in which a prior knowledge of the information signals is available that can be exploited by a DDPM to learn the ``denoising'' process.  
\textcolor{black}{Our scheme is not simply a direct application of the seminal papers in  diffusion model research.  We do not directly apply the seminal DDPM paper \cite{DM_Ho} into a  communication system.    
Rather, different from \cite{DM_Ho}, we incorporate the wireless ``conditions'' into the diffusion model via  employing a \emph{wireless CDiff}.}

The proposed DDPM-assisted  receiver   is employed  in conjunction with the NVIDIA  Sionna simulator\footnote{\url{https://github.com/NVlabs/sionna}}, {by plugging the trained diffusion model to  the receiver side.}   
Data communication pipeline of our scheme is shown in Fig. \ref{fig:pipeline},  which highlights   the compatibility of the scheme with a practical communication system.  
Moreover, in contrast to \cite{INN, hybrid}, the proposed scheme does not rely on any additional autoencoder model.      
Numerical results highlight that the \emph{robustness}  of the wireless system  can be improved 
against  practical non-idealities  such as hardware impairments and channel distortions.  
We compare the reconstruction  performance of our scheme to the conventional digital communications without diffusions, and also the deep neural network (DNN)-based benchmark \cite{DNN}. It is also shown that more than 10 dB improvement in the reconstruction could be achieved in low SNR regimes, 
without the need to reduce  the information rate for error correction. 


\subsection{{Paper Organization and Notations}}
The rest of the paper is organized as follows. In Section \ref{sec:DDPM}, we first  introduce the concept of DDPMs, the forward, and the reverse diffusion processes, together with the main formulas and the corresponding loss functions. 
We also try to point out insightful  intuitions behind the main formulas of the DDPM framework.  
Our system model is introduced in Section \ref{sec:SysMod}. We first  provide the generic formulation of a communication system  under RF  impairments.   We then address the proposed conditional DDPM by modifying the vanilla diffusion model.  Numerical experiments are provided in Section \ref{sec:Eval}. Finally,  Section \ref{sec:concl} concludes the paper. 

\subsubsection*{Notations} 
Vectors and matrices are represented, respectively,  by bold lower-case and upper-case symbols.   Absolute value of a scalar variable and the $\ell_2$ norm of a vector are denoted, respectively by $|\cdot|$ and $||\cdot ||$. Notation $\mathcal{N}(\boldsymbol{x}; \boldsymbol{\mu}, \boldsymbol{\Sigma})$  stands for the multivariate normal distribution  with mean vector $\boldsymbol{\mu}$ and covariance matrix $\boldsymbol{\Sigma}$ for a random vector $\boldsymbol{x}$. Similarly, complex normal distribution with the corresponding mean vector  and covariance matrix is denoted by $\mathcal{CN}(\boldsymbol{\mu}, \boldsymbol{\Sigma})$. Moreover, the expected value of a random variable is denoted by $\mathbb{E}\left[\cdot\right]$   Sets are denoted by calligraphic symbols, and  $\bm 0$ and $\bf I$  show all-zero vector and identity matrix of the corresponding size, respectively. Moreover, $[N]$,  (with $N$ as integer) denotes the set of all integer values from $1$ to $N$, and $\mathsf{Unif}[N]$, $N > 1$, denotes discrete uniform distribution  with samples between $1$ to $N$.  Real and imaginary parts of a complex-valued vector are denoted by $\Re\{\cdot\}$ and $\Im\{\cdot\}$, respectively.

\section{\textcolor{black}{Investigation of Denoising Diffusion Probabilistic  Models}}\label{sec:DDPM} 
\textcolor{black}{In this section, we provide  insights into the underlying mechanisms of the DDPM model. We also provide a comprehensive overview of the theoretical foundations  and practical implications of the DDPM model, offering a deeper understanding of its potential impact on communication systems. 
}

\begin{figure}
	\vspace{0mm}
	\centering
	\includegraphics
	[width=3.3in,height=3.0in,trim={0.0in 0.0in 0.0in  0.0in},clip]{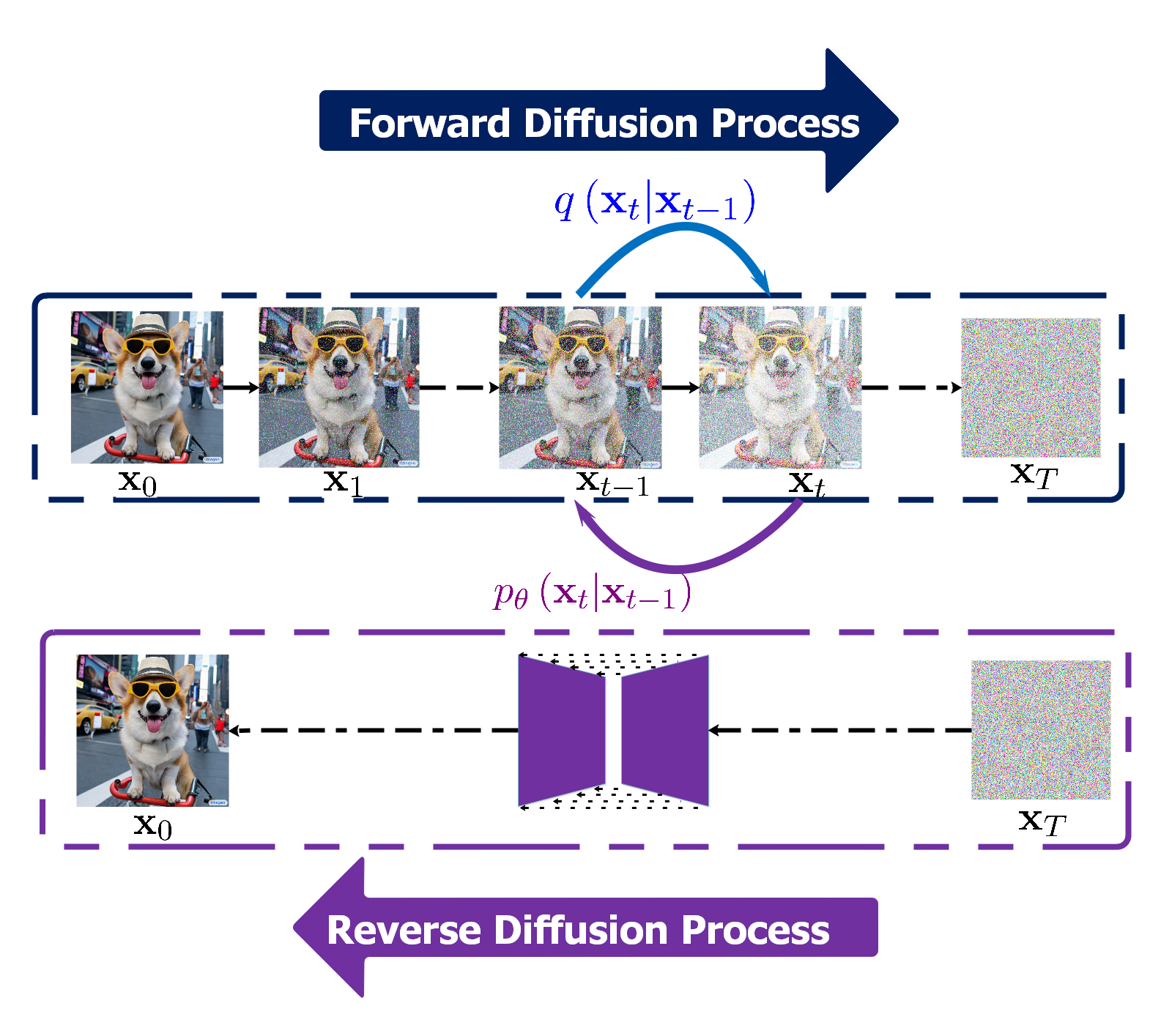}
	\vspace{-1mm}\caption{\small Diffusion models: A general overview.  A Markov chain is defined to mimic the forward diffusion steps, during which random perturbation  noise is purposefully added to the original data. Then in a reverse process, the model learns to construct the desired data samples out of noise. 
 }
	\label{fig:DM}
 \vspace{-2mm}
\end{figure} 

\subsection{\textcolor{black}{Theoretical Foundations}} 

Diffusion models are 
comprised of  two processes, namely  the forward diffusion process and the
parametric reverse diffusion process, as illustrated  in  Fig. \ref{fig:DM}.  
During the forward process, 
 samples are diffused using  Gaussian kernels to add noise until the fully-distorted signals follow an isotropic  Gaussian  distribution. 
Then in the reverse process, the model  tries to ``decode'' the data via ``denoising'' the perturbed samples  in a  hierarchical fashion.  

\subsubsection{Forward Diffusion Process}
 Let ${\bm x}_0$  be a data sample from some  distribution ${ q}({\bm x}_0)$, i.e.,  ${\bm x}_0\sim{q}({\bm x}_0)$. For a finite number, $T$, of time-steps, the forward diffusion process
 $ q({\bm x}_t|{\bm x}_{t-1})$ 
 is defined by adding  Gaussian noise at each time-step 
$t \in [T]$  according to a known ``variance schedule'' $0 < \beta_1  < \cdots <  \beta_T < 1$. 
 This is formulated as 
\begin{align}
    q(\boldsymbol{x}_t \vert \boldsymbol{x}_{t-1}) 
    & \sim \mathcal{N}(\boldsymbol{x}_t; \sqrt{1 - \beta_t} \boldsymbol{x}_{t-1}, \beta_t\boldsymbol{I}),  \label{eq:fwd_diffusion}  
    \\
q(\boldsymbol{x}_{1:T} \vert \boldsymbol{x}_0) 
& = \prod^T_{t=1} q(\boldsymbol{x}_t \vert \boldsymbol{x}_{t-1}). \label{eq:diffusion_eqn}
\end{align}
Invoking \eqref{eq:diffusion_eqn}, the data sample  gradually loses its distinguishable features as the time-step goes on, where with $T\!\rightarrow \!\infty$, ${\bm x}_T$ approaches an isotropic Gaussian distribution.  
According to \eqref{eq:fwd_diffusion},  each new sample at time-step  $t$ can be drawn from a conditional Gaussian distribution with  mean vector  ${\boldsymbol \mu}_t = \sqrt{1 - \beta_t} \boldsymbol{x}_{t-1}$ and covariance matrix ${\bf \Sigma}^2_t = \beta_t \bf I$. Hence, the forward process is realized  by sampling from a Gaussian noise  $\bm{\epsilon}_{t-1} \sim {\cal N}(\bf 0, I)$  and setting 
\begin{align}\label{eq:fwd_sample_gen_diffusion}
	{\bm x}_t = \sqrt{1-\beta_t}&{\bm x}_{t-1} +\sqrt{\beta_t} {\bm{\epsilon}}_{t-1}.
\end{align}
By recursively applying the reparameterization trick from  ML literature \cite{reparam_ML}, we can sample ${\bm{x}}_t$ 
at any arbitrary time-step $t$  in a closed-form expression.  
This results in 
\begin{align} 
\boldsymbol{x}_t  &= \sqrt{\bar{\alpha}_t}\boldsymbol{x}_0 + \sqrt{1 - \bar{\alpha}_t}\bm{\epsilon}_0,  \label{eq:xt_vs_x0} \\ 
    q({\bm x}_t|{\bm x}_0)&\sim\mathcal{N}\left({\bm x}_t;\sqrt{\bar{\alpha}_t}{\bm x}_0,(1-\bar{\alpha}_t)\boldsymbol{I}\right),\label{eq:xt_vs_x0_dist}
\end{align} 
where  
$\bar{\alpha}_t  =\!\prod_{i=1}^t(1-\alpha_i)$  and $\alpha_t=1-\beta_t$.
Note that  \eqref{eq:xt_vs_x0} implies that  we can directly  sample ${\boldsymbol{x}}_t$  from $x_0$ for all $t \in [T]$. 

\subsubsection{Parametric Reverse Diffusion  Process}
Now the problem is to  reverse the process in \eqref{eq:xt_vs_x0} and sample from 
$q(\boldsymbol{x}_{t-1} \vert \boldsymbol{x}_t)$, so that we regenerate  the true samples from  some Gaussian noise  
$\boldsymbol{x}_T$.  
However,  we cannot easily estimate the distribution, since  it requires knowing the distribution of all possible data samples (or equivalently exploiting  the entire dataset).  
Hence,  to 
approximate the conditional probabilities and  run the reverse diffusion process, we need to learn a probabilistic  model $p_{\bm \theta}(\boldsymbol{x}_{t-1} \vert \boldsymbol{x}_t)$ that  is parameterized by ${\bm \theta}$.   
Accordingly, the following expressions can be written  
\begin{align} 
p_{\bm \theta}(\boldsymbol{x}_{t-1} \vert \boldsymbol{x}_t) &\sim \mathcal{N}(\boldsymbol{x}_{t-1}; \boldsymbol{\mu}_{\bm \theta}(\boldsymbol{x}_t, t), \boldsymbol{\Sigma}_{\bm \theta}(\boldsymbol{x}_t, t)).   \label{eq:rev_proc_dist_conditional} 
\end{align}
Now the problem simplifies  to learning the mean vector  $\boldsymbol{\mu}_{\bm \theta}(x_t,t)$ and the covariance matrix  $\boldsymbol{\Sigma}_{\bm \theta}(x_t,t)$ 
for the  probabilistic model $p_{\bm \theta}(\cdot)$, where a neural network (parameterized by ${\bm \theta}$) can be trained to approximate (learn) the reverse process.
When we have  ${\bm x}_0$ as a reference, we can take a small step backwards from noise to generate the data samples, and  the reverse step 
would be formulated as $q({\bm x}_{t-1}|{\bm x}_t,{\bm x}_0)$.  
Mathematically speaking, by knowing the conditional probabilities of  $q({\bm x}_t|{\bm x}_0)$ and $q({\bm x}_{t-1}|{\bm x}_0)$, and utilizing  Bayes rule, we can  derive $q({\bm x}_{t-1}|{\bm x}_t,{\bm x}_0)$ 
with a  similar expression 
to \eqref{eq:rev_proc_dist_conditional}:  
\begin{align}
    q(\boldsymbol{x}_{t-1} \vert \boldsymbol{x}_t, \boldsymbol{x}_0) &\sim \mathcal{N}(\boldsymbol{x}_{t-1}; \hspace{1.5mm} {\tilde{\boldsymbol{\mu}}}(\boldsymbol{x}_t, \boldsymbol{x}_0, t), {\tilde{\beta}_t} \boldsymbol{I}),  
 \label{eq:rev_conditioned_on_x0} \\  
    {\tilde{\boldsymbol{\mu}}}(\boldsymbol{x}_t, \boldsymbol{x}_0, t)
    &=\frac{\sqrt{\alpha_t}(1-\bar{\alpha}_{t-1})}{1-\bar{\alpha}_t}{\bm x}_t + \frac{\sqrt{\bar{\alpha}_{t-1}}
    \beta_t 
    }{1-\bar{\alpha}_t}{\bm x}_0, 
    \label{eq:mu_tilde}  \\ 
   {\tilde{\beta}_t} &=\frac{
   1-\bar{\alpha}_{t-1}}{1-\bar{\alpha}_t} \beta_t.\label{eq:beta_tilde}
\end{align}
Invoking  
\eqref{eq:beta_tilde}, one can infer that the covariance matrix 
has no learnable parameter. 
Hence, we simply need to learn the  mean vector ${\tilde{\boldsymbol{\mu}}}(\boldsymbol{x}_t, \boldsymbol{x}_0, t)$. 
Using the reparameterization trick and with a similar approach to  \eqref{eq:xt_vs_x0},  we can express  ${\bm x}_0$ as 
\begin{align}
    \boldsymbol{x}_0 = \frac{1}{\sqrt{\bar{\alpha}_t}}(\boldsymbol{x}_t - \sqrt{1 - \bar{\alpha}_t}\bm{\epsilon}_t). \label{eq:x0_vs_xt}
\end{align}
Substituting ${\bm x}_0$ in \eqref{eq:mu_tilde} by \eqref{eq:x0_vs_xt} results in 
\begin{align}
    \begin{aligned}
\tilde{\boldsymbol{\mu}}(\boldsymbol{x}_t, \boldsymbol{x}_0, t) = 
{\frac{1}{\sqrt{\alpha_t}} \Big( \boldsymbol{x}_t - \frac{1 - \alpha_t}{\sqrt{1 - \bar{\alpha}_t}} \bm{\epsilon}_t \Big)}.
\end{aligned}
\end{align}
Now we can learn the conditioned probability distribution $p_{\bm \theta}(\boldsymbol{x}_{t-1} \vert \boldsymbol{x}_t)$
of the reverse diffusion process 
by training a neural network   that approximates $\tilde{\boldsymbol{\mu}}(\boldsymbol{x}_t, \boldsymbol{x}_0, t)$.    
Therefore,  we simply need to set the approximated mean vector  $\boldsymbol{\mu}_{\bm \theta}(\boldsymbol{x}_t, t)$ to have the same mathematical format as the target mean vector $\tilde{\boldsymbol{\mu}}(\boldsymbol{x}_t, \boldsymbol{x}_0, t)$.   Hence we have 
 \begin{align}\label{eq:mu_theta_to_learn}
\boldsymbol{\mu}_{\bm \theta}(\boldsymbol{x}_t, t) &=  {\frac{1}{\sqrt{\alpha_t}} \Big( \boldsymbol{x}_t - \frac{1 - \alpha_t}{\sqrt{1 - \bar{\alpha}_t}} \bm{\epsilon}_{\bm \theta}(\boldsymbol{x}_t, t) \Big)}, 
\end{align}
where $\bm{\epsilon}_{\bm \theta}(\boldsymbol{x}_t, t)$ denotes the model to predict $\bm{\epsilon}_t$.   

We now define the loss function $\mathcal{L}_t$ for time-step $t$, aiming  to minimize the difference between $\boldsymbol{\mu}_{\bm \theta}(\boldsymbol{x}_t, t)$ and  $\tilde{\boldsymbol{\mu}}(\boldsymbol{x}_t, \boldsymbol{x}_0, t)$.   
\begin{align}
\mathcal{L}_t &=  
{\mathbb{E}}_{\begin{subarray}{l}t\sim {\mathsf{Unif}}[T]\\ {\boldsymbol{x}}_0\sim q({\boldsymbol{x}}_0) \\ \bm{\epsilon}_0\sim \mathcal{N}(0,\boldsymbol{I})\\ \end{subarray}}
\Big[\|\bm{\epsilon}_t - \bm{\epsilon}_{\bm \theta}(\boldsymbol{x}_t, t)\|^2 \Big]  \nonumber \\  
&=  {\mathbb{E}}_{\begin{subarray}{l}t\sim {\mathsf{Unif}}[T]\\ {\boldsymbol{x}}_0\sim q({\boldsymbol{x}}_0) \\ \bm{\epsilon}_0\sim \mathcal{N}(0,\boldsymbol{I})\\ \end{subarray}}  \Big[\|\bm{\epsilon}_t - \bm{\epsilon}_{\bm \theta}(\sqrt{\bar{\alpha}_t}\boldsymbol{x}_0 + \sqrt{1 - \bar{\alpha}_t}\bm{\epsilon}_t, t)\|^2 \Big]. \label{eq:loss_func}
\end{align}
Invoking 
\eqref{eq:loss_func}, 
at each time-step $t$,  the  DDPM model 
takes $\boldsymbol{x}_t$ as input {and returns the distortion components $\bm{\epsilon}_{\bm \theta}(\boldsymbol{x}_t,t)$.} Moreover, $\bm{\epsilon}_t$ denotes the  diffused noise term  at time-step $t$.     

\vspace{-3mm}

\textcolor{black}{ 
\subsection{Practical Implications and Potential Impacts on Communication Systems}
As discussed above, DDPMs are inherently connected to the task of \emph{denoising}. This makes them highly relevant to communication systems, where the  unwanted noise and interference is a significant challenge. 
This implies some of the potential impacts of DDPMs on  communication systems.  
Notably, DDPMs as  powerful denoising models can be applied to clean noisy signals received over communication channels. The reverse diffusion process could help  reconstruct original signals from noisy transmissions, leading to  improved  signal quality \cite{wcl}. 
Further, DDPMs could be integrated into an end-to-end  framework, where both the transmitter and receiver are learned jointly. The transmitter could use a forward diffusion process to generate noisy encoding of the transmitted signal, while the receiver could use the reverse process to decode the original data, optimizing the overall communication pipeline for robustness and efficiency \cite{DM_for_E2EComm}. 
}
\textcolor{black}{
To conclude, we note that theoretical robustness and practical applications of DDPMs, introduced in this section, highlight  that they have the potential to play important roles in future communication systems. This can include, but is not limited to, \emph{resilient communications}, making communication systems more resilient to noise, interference, and distortions, particularly in extreme environments,  such as space (non-terrestrial networks), underwater, or urban areas with heavy interference. High-fidelity signal reconstruction is another direct impact of employing  DDPMs for communication networks, which can potentially result in clearer voice calls, sharper video transmissions, and more accurate data recovery in practical communication applications. 
}

\section{System Model and Proposed Scheme}\label{sec:SysMod}  
\subsection{System Model \& Formulas}\label{subsec:ProblemFormula}  
Consider a point-to-point communication system, where a source node (the transmitter)   
aims to transmit the ground-truth information $x_0 \sim q(x_0)$ to a destination node (the receiver) over the communication channel.  
We denote by $s$, the 
channel input signal subject to the power constraint $\mathbb{E}[|s|^2]\leq P$, for which we can write $s = g(x_0)$ with $g(\cdot)$ denoting the mapping from information signals to the channel input (including the off-the-shelf  modulation and coding functionalities). The information-bearing signal $s$ is then sent over the communication channel and $y = \eta(s)$ is observed  at the receiver, where $\eta(\cdot)$ stands for the channel distortion and noise, as well as  the impairments caused by the non-ideal transceiver  hardware.  
To model $\eta(\cdot)$, we consider a practical scenario of hardware-impaired communication, where we employ the well-known  generic 
 communication  model of \cite{HI}. 
Denoting the Rayleigh fading wireless channel coefficient 
by $h$,  
we have 
\begin{align}\label{eq:HWI_scalar}
{y} &= {h} (\sqrt{P} {s} + {\tau}^{t}) + \tau_{}^{r} + n,  \\
\tau_{}^t &\sim \mathcal{CN}({0},\kappa^t P), \quad \tau^r\sim \mathcal{CN}(0,\kappa^r P|h_{}|^2), 
\end{align}
where $P$ denotes the transmit power,  
$\tau_{}^t$
is the distortion noise caused by the transmitter hardware with the corresponding impairment level $\kappa^t$, and   
$\tau_{}^{r}$ reflects the hardware distortion at the receiver with $\kappa^r$ showing the level of impairment at the 
receiver hardware.  
Notably, ${\tau^r}$ is conditionally Gaussian, given the channel realization  ${h}$. 
{This is a well-known and  experimentally-validated model for HIs, which is widely-adopted in wireless communication literature  \cite{HI}.} 
We emphasize that  according to \cite{HI},  
the distortion noise caused at each radio frequency (RF) device  is proportional to its signal power, where    
a fixed portion of the information signal is turned into \emph{distortion noise} due to 
quantization errors in analog-to-digital converter (ADC), 
inter-carrier interference induced by phase noise, leakage from the mirror sub-carrier under I/Q imbalance, non-linearities in power amplifiers, etc.  
To be aligned with the ``batch-processing'' nature of AI/ML algorithms and wireless communications  simulators  \cite{Sionna}, we  reformulate the signaling expressions in matrix-based format.
To do so, we define the  batch (with size $K$) of information-bearing  samples as ${\boldsymbol{s}} \overset{\Delta}{=}  \left[s_1,  s_2, \cdots, s_K\right]^{\sf T}$ corresponding to the batch of ground-truth information samples ${\boldsymbol{x}_0} \overset{\Delta}{=}  \left[x_0^{(1)},  x_0^{(2)}, \cdots, x_0^{(K)}\right]^{\sf T}$ with the underlying distribution $\boldsymbol{x}_0 \sim q(\boldsymbol{x}_0)$.   Hence, the  batch
of 
received signals at the receiver  can be expressed as          
\begin{align}\label{eq:received_y_tensor}
    \boldsymbol{y} =  \sqrt{P} \hspace{1mm} \boldsymbol{H}  \boldsymbol{s} + \bm{\zeta}, 
\end{align}
where $\boldsymbol{H} =  \hspace{1mm} \mathsf{diag}(\boldsymbol{h})$, ${\boldsymbol{h}} \overset{\Delta}{=} \left[ h^{(1)},  h^{(2)}, \ldots, h^{(K)} \right]^{\sf T}$ stands for  the  channel realizations vector corresponding to the transmission of information samples, and   ${\bm \zeta} \overset{\Delta}{=}  \boldsymbol{H}\bm{\tau}^t  + \bm{\tau}^r  + \boldsymbol{n}$ represents  the  
effective noise-plus-distortion at the receiver with the conditional distribution $\bm{\zeta} \vert{\boldsymbol{h}} \sim \mathcal{CN}({{\bf 0}}_K,{\boldsymbol{\Sigma}})$ and  
the covariance matrix ${\bf \Sigma}$ given by
${\bf \Sigma} = P (\kappa^t + \kappa^r) \boldsymbol{G} + \sigma^2 \boldsymbol{I}_K$.
We also define  
\begin{align}
{\bm \tau}^t &\overset{\Delta}{=} \left[  \tau^t_1, \tau^t_2,  \ldots,  \tau^t_K \right]^{\sf T}
\sim \mathcal{CN}({{\bf{0}}_K},\kappa^t P {\boldsymbol{I}}_K), \\ 
    {\bm{\tau}}^r &\overset{\Delta}{=} \left[  \tau^r_1,  \tau^r_2, \ldots,  \tau^r_K \right]^{\sf T} 
    \sim \mathcal{CN}({\bf 0}_K, \kappa^r P {\boldsymbol{G}}), 
\end{align}
which  is conditionally Gaussian given the channel realizations  $\{{h}_k\}_{k \in [K]}$,   with $\boldsymbol{G} = \mathsf{diag}(\left[ |h_1|^2, |h_2|^2, \ldots, |h_K|^2 \right]^{\sf T})$. 
Due to the fact that neural networks can only process real-valued inputs,  
we  map complex-valued symbols to real-valued tensors, and rewrite 
\eqref{eq:received_y_tensor}
by stacking the real and imaginary components. This results in the following expression.   
\begin{align} \label{eq:received_z_tensor} 
\tilde{\boldsymbol{y}}_{} = \tilde{\boldsymbol{H}} \tilde{\boldsymbol{s}} + \bm{\nu}, 
\end{align}
where 
\begin{align}
\tilde{\boldsymbol{y}}_{}&=\begin{bmatrix} \Re\{{\boldsymbol{y}}\} \\ \Im\{{\boldsymbol{y}}\} \end{bmatrix} \in \mathbb{R}^{2K}, 
\\ 
\tilde{{\boldsymbol{s}}} &= \sqrt{p} \begin{bmatrix}  \Re\{{\boldsymbol{s}}\} \\ \Im\{{\boldsymbol{s}}\} \end{bmatrix} \in \mathbb{R}^{2K} \label{eq:z-x}, 
\\
{\tilde{{\boldsymbol{H}}}}&=\begin{bmatrix} \Re\{{\boldsymbol{H}}\} & -\Im\{{\boldsymbol{H}}\} \\ \Im\{{\boldsymbol{H}}\} & \Re\{{\boldsymbol{H}}\}\end{bmatrix} \hspace{-0.25mm} \in   \hspace{-0.25mm} \mathbb{R}^{2K \times 2K},
\\ {\bm \nu} &= \begin{bmatrix} \Re\{\bm{\zeta}\} \\ \Im\{\bm{\zeta}\} \end{bmatrix} \hspace{-0.5mm} \in   \hspace{-0.5mm} \mathbb{R}^{2K} \label{eq:H-v}.
\end{align}


The receiver tries to obtain the information signals, using a decoding functionality $f(\cdot)$, which reverts the mapping carried out at the transmitter,  either via conventional digital communication-based  methods, or the DNN-based algorithms as in \cite{DNN}. Accordingly, a noisy  version of the ground-truth batch, denoted by $\widehat{\boldsymbol{x}}$ would  be obtained 
as 
\begin{align}
\widehat{\boldsymbol{x}} = f(\eta(g(\boldsymbol{x}))) \approx  \boldsymbol{x}_0 + \boldsymbol{z},  
\label{eq:x_hat_x_z}
\end{align}
where $\boldsymbol{z}$ denotes the  unknown 
reconstruction error.  

\begin{remark}
\it 
Notably, in extreme conditions due to low SNR regimes or highly mismatched transceiver and receiver due to large impairment levels, (where even error correction capabilities might not be satisfactory), the idea is to leverage the ``denoising-and-generation'' characteristics of DDPMs to help facilitate obtaining the ground-truth information signals.  
\end{remark}

\textcolor{black}{
\begin{remark}
\it 
The vanilla format of denoising process of DDPMs,  as introduced in Section \ref{sec:DDPM}, assumes starting the denoising-and-generation process with an isotropic Gaussian noise.   
However, in a practical wireless
system, the reconstructed signals  at the receiver are a degraded version, $\widehat{\bm x}$, of the ground-truth information signals, not necessarily a pure 
isotropic Gaussian noise. This degradation is different  in different distortion conditions such as different SNRs
and impairment levels.  
\end{remark}
To address the above-mentioned  challenges,  we incorporate  the noisy 
degraded signals into the denoising framework of the diffusion model, realizing a conditional DDPM. 
Accordingly, the denoising process is modified and the
corresponding conditional diffusion and reverse processes are derived.   
}

\begin{figure}
	\vspace{0mm}
	\centering
	\includegraphics
	[width=3.45in,height=2.3in,trim={0.0in 0.0in 0.0in  0.0in},clip]{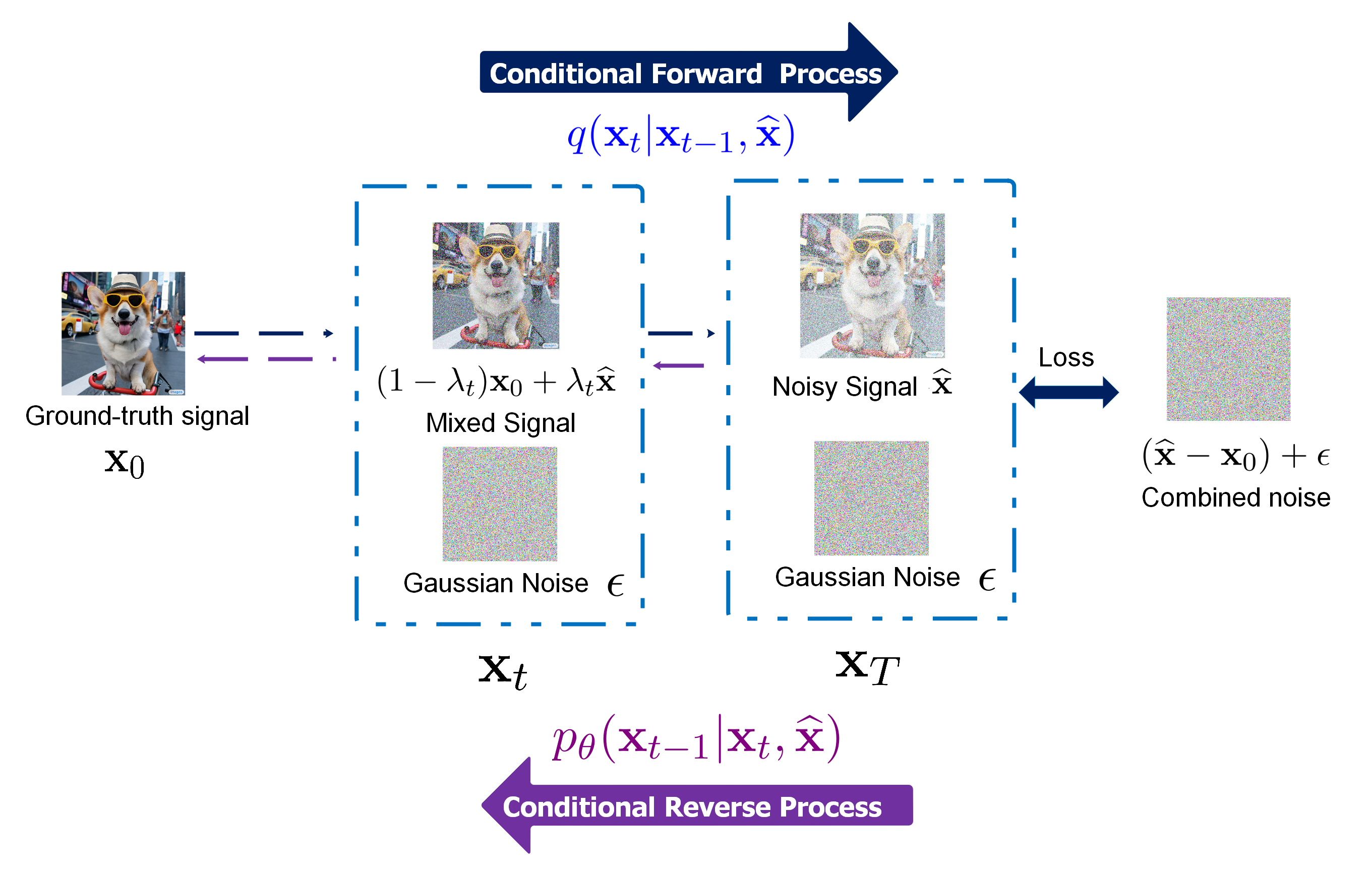}
	\vspace{-1mm}\caption{\small Schematic of the Conditional DDPM Framework. 
 }
	\label{fig:CDiff}
 \vspace{-2mm}
\end{figure}

\subsection{Proposed Solution: CDiff  for Enhanced  Data Reconstruction}\label{subsec:Algorithms} 

\subsubsection{Conditional Forward Process}
The idea  of the employed  conditional DDPM is to first incorporate the noisy degraded  samples into the forward diffusion framework, so that we can learn,  in the reverse diffusion, to remove the exact noise from $\widehat{\bm x}$ and refine the ground-truth information  signals.    
For this,  a weighted sum of 
the clean version of ground-truth samples ${\bm x}_0$ and the noisy samples $\widehat{\bm x}$ are considered, using  an interpolation hyperparameter $\lambda_t$, as shown in Fig. \ref{fig:CDiff}.  
Then the Markov chain-like transition model $q(x_t|x_{t-1})$ in \eqref{eq:fwd_diffusion} is modified into a  conditional forward  diffusion process as follows.  
\begin{align} 
  q_{\sf{cond}}({\bm x}_t|{\bm x}_0,{\widehat{\bm x}}) \sim  \mathcal{N}\Big({\bm x}_t;(1-\lambda_t)\sqrt{\bar{\alpha}_t}{\bm x}_0 + \lambda_t\sqrt{\bar{\alpha}_t}\widehat{\bm x} ,\delta_tI\Big),
  \label{eq:new diffusion process t}
\end{align} 
Invoking \eqref{eq:new diffusion process t}, one can observe that  this is a generalization of the vanilla  forward diffusion process in 
 \eqref{eq:xt_vs_x0_dist}, as we assume that the mean vector  reflects an interpolation between the clean information signals ${\bm x}_0$ and their noisy degraded version $\widehat{\bm x}$. 
Accordingly, $\lambda_t$ is set to start from $0$ and  gradually increased to $\lambda_T \approx 1$. Intuitively, this guides the diffusion model to pay more attention to learning the denoising process from noisy samples over time. 
According to \eqref{eq:new diffusion process t}, one can marginalize the conditional distributions over $\widehat{\bm x}$ and write 
\begin{align}
    q_{\sf{cond}}({\bm x}_t|{\bm x}_0)=\int  \hspace{-0.5mm} q_{\sf{cond}}({\bm x}_t|{\bm x}_0,\widehat{\bm x}) \hspace{0.7mm} p_{\widehat{\bm x}}(\widehat{\bm x}|{\bm x}_0) \hspace{0.7mm} {d}\widehat{\bm x}.  
    \label{eq:integral_format}
\end{align} 
In order for the conditional diffusion $ q_{\sf{cond}}({\bm x}_t|{\bm x}_0)$ to become a  generalization  of the vanilla  diffusion process $q({\bm x}_t|{\bm x}_0)$ in \eqref{eq:xt_vs_x0_dist},  one needs to set 
\begin{align}
    \delta_t := (1-\bar{\alpha}_t)-\lambda_t^2\bar{\alpha}_t. 
  \label{eq:new diffusion process t delta}
\end{align} 

\begin{algorithm}[!tbp]
\hspace*{0.02in} {\bf {Hyper-parameters:}}
	{Number of diffusion steps $T$, neural architecture $\boldsymbol{\epsilon}_{\theta}(\cdot, \cdot, t)$, variance schedule   $\beta_t$ and $\bar{\alpha}_t,$ and conditioning weights $\lambda_t$,  $\forall t \in [T]$.} \\
    \hspace*{0.02in} {\bf {Input:}}
	{Training samples: The pair of ground-truth information signal and its noisy decoded version from a dataset $\cal D$: $(\boldsymbol{x}_0, {\widehat{\bm x}}) \in \mathcal{D}$} \\
	\hspace*{0.02in} {\bf {Output:}} {Trained model $\boldsymbol{\epsilon}_{\theta}(\cdot, \cdot, t)$.}
	\caption{Training algorithm}
	\label{alg:trainAlg}
	\begin{algorithmic}[1] 
    \WHILE {the stopping criteria are not met}
    \STATE Randomly sample $(\boldsymbol{x}_0, {\widehat{\bm x}})$ from $\cal D$ 
    \STATE Randomly sample $t \sim \mathsf{Unif}[T]$ 
    \STATE Randomly sample $\boldsymbol{\epsilon} \sim \mathcal{N}(\boldsymbol{0},\boldsymbol{I})$ 
    \STATE Set $\boldsymbol{x}_t = (1-\lambda_t)\sqrt{\bar{\alpha}_t}{\bm x}_0 +  \lambda_t\sqrt{\bar{\alpha}_t}\widehat{\bm x} + \delta_t \boldsymbol{\epsilon}$
      \STATE Take gradient descent step on
      \STATE $\quad {\grad}_{\boldsymbol{\theta}} \left\| 
      \frac{1}{\sqrt{1-\bar{\alpha}_t}}
({\lambda_t\sqrt{\bar{\alpha}_t}}{({\widehat{\bm x}}-{\bm x}_0)} + {\sqrt{\delta_t}}{\bm\epsilon}) - {\bm \epsilon}_\theta({\bm x}_t, \widehat{\bm x}, t) 
      \right\|^2$
    \ENDWHILE
	\end{algorithmic}
\end{algorithm}

\subsubsection{\textcolor{black}{Conditional Reverse Process}} 
\label{Sec:CDiffuSE reverse}
\begin{figure}
	\vspace{0mm}
	\centering
	\includegraphics
	[width=3.5in,height=2.2in,trim={0.0in 0.0in 0.0in  0.0in},clip]{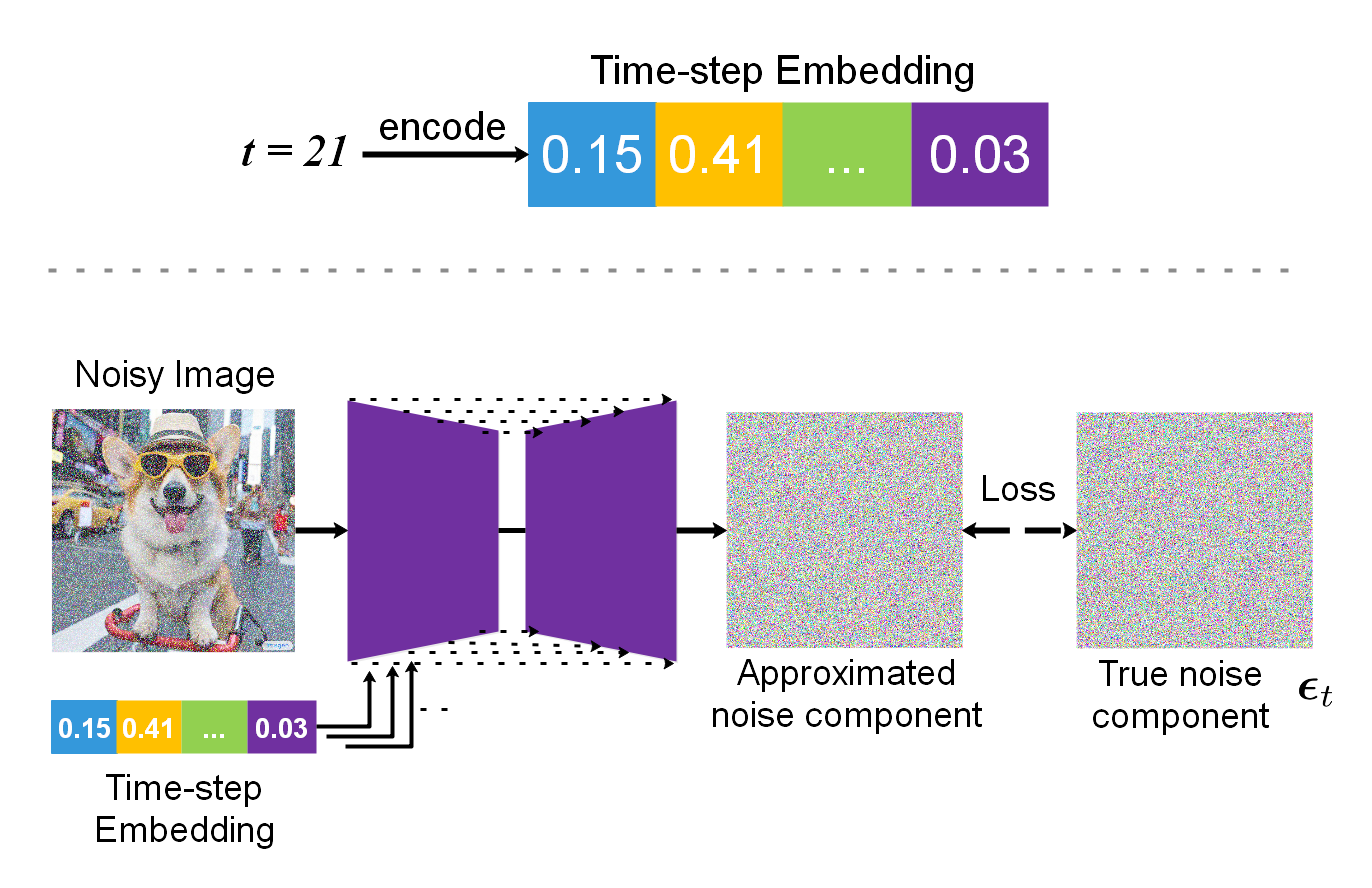}
	\vspace{0mm}\caption{\small General  description of training the diffusion model \textcolor{black}{and the notion of time embedding.   Each time-step is mapped (via positional embedding) to an embedding vector, which is  then incorporated into the hidden layers of the neural network. The figure shows time-step $t=21$ as an example.}}
	\label{fig:embeddings}
 \vspace{0mm}
\end{figure}

The reverse process starts from ${\bm x}_T$ as the degraded information signal $\widehat{\bm x}$ 
according to \eqref{eq:new diffusion process t} while setting  $\lambda_T = 1$, where we have 
$  p_{\sf{cond}}({\bm x}_{T}|\widehat{\bm x}) \sim  \mathcal{N}({\bm x}_T,\sqrt{\bar{\alpha}_T}\widehat{\bm x},\delta_T {\bm I}).$  
Considering a similar approach to \eqref{eq:rev_proc_dist_conditional}, the conditional reverse process 
is considered to predict ${\bm x}_{t-1}$ based on ${\bm x}_{t}$ and $\widehat{\bm x}$. This results in the following conditional reverse transition probability distribution.  
\begin{equation}
  p_{\sf{cond}}({\bm x}_{t-1}|{\bm x}_t, \widehat{\bm x}) \sim  \mathcal{N}({\bm x}_{t-1};{\bm \mu}_\theta({\bm x}_t,\widehat{\bm x},t), {\delta}_t {\bm I}),
  \label{eq:new reverse}
\end{equation} 
with ${\bm \mu}_\theta({\bm x}_t,\widehat{\bm x},t)$ formulating the estimated mean of the conditional reverse process. 
The expression for the conditional mean  of the reverse process in \eqref{eq:new reverse} should be a generalization to the vanilla format,  
by incorporating the noisy signals $\widehat{\bm x}$ into the formulation as well.  Therefore, with a similar expression  to \eqref{eq:mu_theta_to_learn}, the mean vector for each step  of the conditional denoising can be expressed  as a weighted sum of the denoised signal at step $t$, ${\bm x}_t$, the conditional information regarding the degraded ground-truth,  $\widehat{\bm x}$, and the estimated noise $\bm \epsilon_{\theta}(\cdot, t)$. Hence, we have  
\begin{equation}
  {\bm \mu}_\theta({\bm x}_t,\widehat{\bm x},t) = \psi_{x}{\bm x}_{t} + \psi_{\hat{x}} {\widehat{\bm x}} - \psi_{\epsilon} {\bm \epsilon}_\theta({\bm x}_t, \widehat{\bm x}, t),
  \label{eq:new reverse mean}
\end{equation}
It is shown in \cite{cdiff} that the values of coefficients $ \psi_{x}, \psi_{\hat{x}}, $ and $ \psi_{\epsilon}$ can be estimated  by solving the evidence lower bound (ELBO) optimization criterion, where we have

\begin{align}
    \psi_{x} &= \frac{\delta_{t-1} (1-\lambda_t)}{\delta_{t} (1-\lambda_{t-1})}\sqrt{\alpha}_t + (1-\lambda_{t-1})\frac{\delta_{t|t-1}}{\delta_{t}\sqrt{\alpha}_t}, \label{eq:c_{xt}} \\
    \psi_{\hat{x}} &= (\lambda_{t-1}\delta_t - \frac{\lambda_t(1-\lambda_t)}{1-\lambda_{t-1}}\alpha_t\delta_{t-1})\frac{\sqrt{\bar{\alpha}_{t-1}}}{\delta_t}, \label{eq:c_{yt}} \\
    \psi_{\epsilon} &= (1-\lambda_{t-1})\frac{\delta_{t|t-1}\sqrt{1-\bar{\alpha}_t}}{\delta_t \sqrt{\alpha_t}}, 
  \label{eq:c_epsilon}
\end{align}
where $\delta_{t|t-1} \overset{\Delta}{=} \delta_t - \left(\frac{1-\lambda_t}{1-\lambda_{t-1}}\right)^2 \alpha_t \delta_{t-1}$. 
As such, the loss function for the conditional model is expressed as 
\begin{align}
    &\mathcal{L}_t^{\sf cond} = 
    \nonumber \\ 
   & 
    \mathbb{E}_{{\bm x}_0,{\bm \epsilon}, \widehat{\bm x}} \parallel (\frac{\lambda_t\sqrt{\bar{\alpha}_t}}{\sqrt{1-\bar{\alpha}_t}}{({\widehat{\bm x}}-{\bm x}_0)} + \frac{\sqrt{\delta_t}}{\sqrt{1-\bar{\alpha}_t}}{\bm\epsilon}) - {\bm \epsilon}_\theta({\bm x}_t, \widehat{\bm x}, t) \parallel^2_2, 
  \label{eq:ELBO_optimized}
\end{align}
where $\bm \epsilon \sim \mathcal{N}(0, \bm I)$.
Intuitively speaking, the loss function in \eqref{eq:ELBO_optimized} is designed such that the conditional DDPM learns, over the denoising steps, to estimate both the Gaussian noise $\bm \epsilon$ and the residual errors of estimating $\widehat{\bm x}$, i.e.,  ${\widehat{\bm x}}-{\bm x}_0$. \textcolor{black}{Therefore, 
the coefficients} for $({\widehat{\bm x}}-{\bm x}_0)$ and $\bm \epsilon$ are chosen similar to the coefficient of ${\widehat{\bm x}}$ and the standard deviation of the diffusion process as in 
\eqref{eq:new diffusion process t},  respectively.  
According to \eqref{eq:new diffusion process t delta}--\eqref{eq:ELBO_optimized}, the training and sampling algorithms are summarized in Algorithm \ref{alg:trainAlg} and Algorithm \ref{alg:sampling}, respectively.  A visual description of the  training phase is also exhibited in Fig. \ref{fig:embeddings}.

\begin{algorithm}[t!]
  \caption{ Sampling  algorithm} \label{alg:sampling}
  \begin{algorithmic}[1]
    \vspace{0.0in}
    \STATE Sample $\bx_T \sim \mathcal{N}(\bx_T; \sqrt{\bar{\alpha}_T} \widehat{\boldsymbol{x}}, \delta_T \boldsymbol{I})$ 
    \FOR{$t=T, ... , 1$}
      \STATE $\bz \sim \mathcal{N}(\bzero_{}, \bI_{})$ 
      \STATE $\bx_{t-1} = \psi_{x}{\bm x}_{t} + \psi_{\hat{x}} {\widehat{\bm x}} - \psi_{\epsilon} {\bm \epsilon}_\theta({\bm x}_t, \widehat{\bm x}, t) + 
      \sqrt{\delta_t} \bz$
    \ENDFOR 
    \STATE \textbf{return} $\bx_0$
    \vspace{0.0in}
  \end{algorithmic}
\end{algorithm}

\section{Numerical Results}\label{sec:Eval}
In this section, we provide different numerical evaluations  to highlight the performance of our proposed approach in different scenarios and compared to different baselines.   
We also evaluate  the  \emph{robustness} of our proposed  scheme  under  low-SNR regimes.

\textcolor{black}{
\subsection{Experimental Setup}\label{subsec:Setup}}

\subsubsection{\textcolor{black}{\textbf{Communication Setup}}} 
\textcolor{black}{
Our scheme is implemented in conjunction with  the NVIDIA  Sionna simulator \cite{Sionna} 
to show  the compatibility of the scheme with a practical communication system. {To elaborate, after that the diffusion model is trained offline using Algorithm \ref{alg:trainAlg}, it is plugged into the end of the receiver block of the Sionna simulator as shown in Fig. \ref{fig:pipeline}.  
In the evaluations, both scenarios of DNN-based receiver, known as Sionna's neural receiver, and also the conventional receiver are considered in the evaluations.}\footnote{{One interesting direction would be to propose a framework in which the forward and reverse process of diffusion training are directly incorporated into the communication simulator. This requires modeling the conditional  probability distribution of data transmission and  reconstruction as modified forward and reverse diffusion processes, respectively, which will be studied in our future works.}
}    
We consider 64-QAM modulation symbols  with 5G low density parity check coding (LDPC) \cite{Sionna} for the modulation and coding of the transmission scheme.   
The transmit SNR is defined as 
\begin{align}
    \Gamma = 10 \log_{10} (\frac{P}{\sigma^2}) \hspace{2mm}\text{dB}. 
\end{align}
Without loss of generality, we normalize the transmit power by setting the average signal power to $P = 1$ for all experiments, while varying the
SNR by setting the standard deviation (std)  of noise $\sigma$ \cite{AE_paper}. 
Following the experimentally-validated model of \cite{HI},  we consider  the impairment  levels  in the range {$[0,0.15^2]$} \cite{wcl, conf_hwi},     
where smaller values correspond to  less-impaired  transceiver hardware.}

\subsubsection{\textcolor{black}{\textbf{Learning Setup}}} 
Evaluations are carried out on \textcolor{black}{MNIST dataset\footnote{\url{http://yann.lecun.com/exdb/mnist}} {(resized to $32\times 32$)}. The dataset  is split into  60,000 training
images and 10,000 testing images of hand-written digits \cite{mnist}.  The corresponding dataset $\mathcal{D}$ in Algorithm \ref{alg:trainAlg}, is then obtained by transmitting the training samples of MNIST over the Sionna's simulator and obtain the decoded signals at the receiver's output,  and then pairing each of the ground-truth samples   with their degraded version to create $(\boldsymbol{x}_0, {\widehat{\bm x}}) \in \mathcal{D}$. 
For validation of our numerical analysis, mean-squared error (MSE)  and  PSNR metrics are studied in this paper. These metrics measure the quality of reconstruction, where a higher(lower) PSNR(MSE) indicates better quality.}   

Training of our scheme can be carried out via two different training strategies, i.e.,  varying training SNR (simulating the system with different transmit SNRs), or  fixed training SNR (fixing the communication scenario to a determined SNR).      
Our validations in terms of the PSNR  metric showed that the performance of the fixed training SNR scenario saturates too early during the validation phase.    
Therefore, we choose the varying scenario, where we set $-15 \leq \Gamma^{\sf train} \leq 15$ dB  by \textcolor{black}{trial and error} over different ranges of SNRs.  
\textcolor{black}{Notably, 
having training samples $(\boldsymbol{x}_0, {\widehat{\bm x}})$ from  a wide range of communication SNRs and the typical range of transmitter and receiver impairment levels {$ \kappa^{t, r} \in [0, 0.15^2]$} (See Table \ref{table:Commparams}) indicates that  the  wireless conditions are implicitly taken into account as the conditioning of our model. This   way, we incorporate the conditions  of a  practical wireless system into account.       
}

{For the choice of neural network for our CDiff  model, we follow the model employed in \cite{DM_Ho}, and  adopt the U-Net architecture \cite{unet} as our denoising neural network  ${\bm \epsilon}_\theta({\bm x}_t, \widehat{\bm x}, t)$ with input size 32 and depth of 4. The employed neural network is supposed to take the distorted image at each time step,  and output the predicted noise with the same size as the input image.  
The U-Net architecture matches well with this requirement. It consists of a contracting path and an expansive path, which respectively capture contextual information and reduce the spatial resolution of the input, and then 
up-samples the feature maps, while also performing convolutional operations \cite{unet}.    
Additionally,  it introduced residual connections between the encoder and decoder, which greatly improves the  gradient flow.}   
The training is carried out for 10 epochs over $T=200$ steps, and using  adaptive moment estimation (Adam) optimizer with  learning rate  $\lambda = 10^{-3}$.  To stabilize the training process,  exponential moving average (EMA) method  is  implemented \cite{EMA} {for model update during training}. 
This  helps  maintain 
a form of ``model momentum.''  Specifically, instead of directly updating model's weights, a copy of the previous weights is kept, and the weighted average  between the previous and new version of the weights are calculated for the model update.       
 For the conditioning weights $0<\lambda_t<1$, we consider a linear scheduling over  $T=200$ steps. For the variance scheduling $\beta_t$, we also consider a linear scheduling, starting from $\beta_0 = 0.0001$ to $\beta_T = 0.0095$.  
We note that the values for the variance scheduling were obtained based on trial-and-error over different scheduling formats.  
To the best of our knowledge, there does not exist a unified generic  solution for the choices of variance scheduling in the diffusion model literature, and it depends on the specific problem to which the diffusion model is applied.     
To enable  employing only one neural  network for  the entire denoising steps, instead of training $T$ distinct models for each time-step $t \in [T]$,  we share the   parameters of the neural network  across time-steps. More specifically, as shown in Fig. \ref{fig:embeddings},  we encode the time-steps $t \in [T]$ and input it to each hidden layer of our neural network 
as a positional  embedding \cite{DM_Ho}.   
The output of the hidden layers are then  
multiplied by the time embeddings \cite{DM_Ho, DM_survey, CGM_ChanEst}.   
Intuitively, this makes the model ``know''  at which particular time-step it is operating for every sample in the batch.  
\textcolor{black}{To summarize, Tables \ref{table:Commparams} and \ref{table:MLparams} highlight  the experimental setup and hyper-parameters and metrics used for the simulation.}

\begin{table}[tbp!]
\textcolor{black}{  
\caption{
        \textcolor{black}{ Parameters of the communication setup.}  
    }
    \centerline{
\begin{tabular}{|l|l|}
  \hline
  Parameter & Value \\
  \hline
  Simulation Setup & NVIDIA Sionna Link-level Simulator \cite{Sionna} \\
    \hline 
  Modulation scheme & 64 QAM \cite{Sionna}\\
  \hline 
 Coding scheme & 5G LDPC \cite{Sionna} \\
  \hline 
  Training SNR &  {$[-15, 15]$} dB \\
  \hline 
  Impairment level  & {$[0, 0.15^2]$} \cite{HI, wcl, conf_hwi}  \\
  \hline 
\end{tabular}} 
\label{table:Commparams}
}
\end{table}

\begin{table}[tbp!]
 \textcolor{black}{    \caption{
        \textcolor{black}{ Hyper-parameters of the learning  setup.}  
    }
    \centerline{
\begin{tabular}{|l|l|}
  \hline
  Parameter & Value \\
  \hline
  Dataset & MNIST \cite{mnist} \\
    \hline 
  Train-test split & 60000 - 10000 \\
  \hline 
 Neural architecture & U-Net with depth 4 (32, 64, 128, 256) \cite{DM_Ho, unet}  \\
  \hline 
  Total denoising steps &  $T = 200$ \\
  \hline 
  Training epochs  & 10  \\
  \hline 
  variance scheduling & linear for ($\beta_0, \beta_T) = (0.001, 0.0095)$ \cite{DM_Ho} \\ 
  \hline 
  Learning rate  & 0.001 with Adam optimizer  \cite{EMA} 
  \\
  \hline 
\end{tabular} 
\label{table:MLparams}
}}
\end{table} 

\textcolor{black}{\subsection{Baseline methods used for comparison}\label{subsec:baselines}} 
\textcolor{black}{The following baselines are employed for comparison with our proposed approach.} 

\begin{figure*}
\centering
    \begin{subfigure}{0.325\textwidth}
        \centering    
        \includegraphics[width=\linewidth, trim={0.0in 0.0in 0.0in  0.0in},clip]{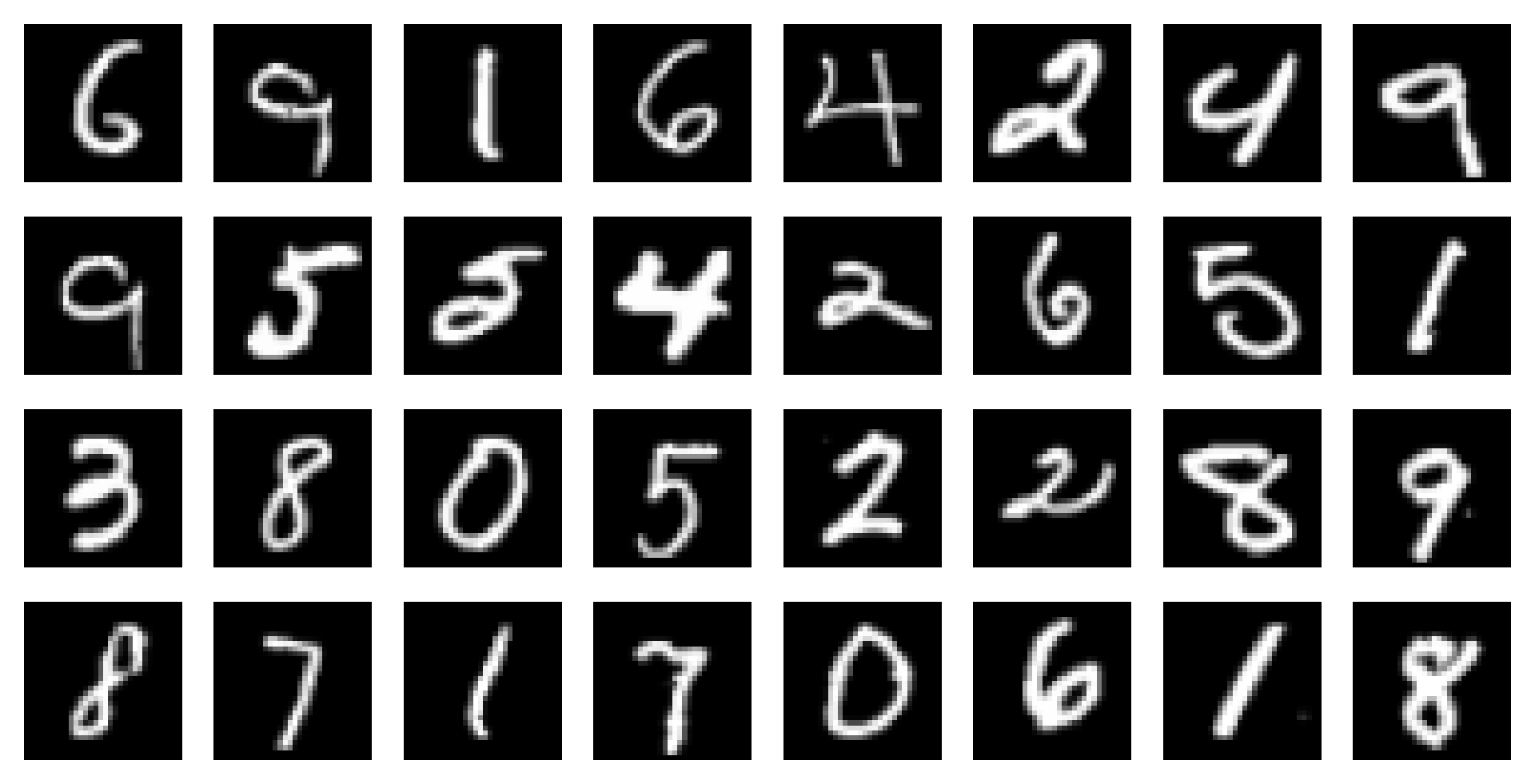}
        \caption{Original images $\boldsymbol{x}_0$}
    \end{subfigure}
     \begin{subfigure}{0.325\textwidth}
        \centering    
        \includegraphics[width=\linewidth, trim={0.0in 0.0in 0.0in  0.0in},clip]{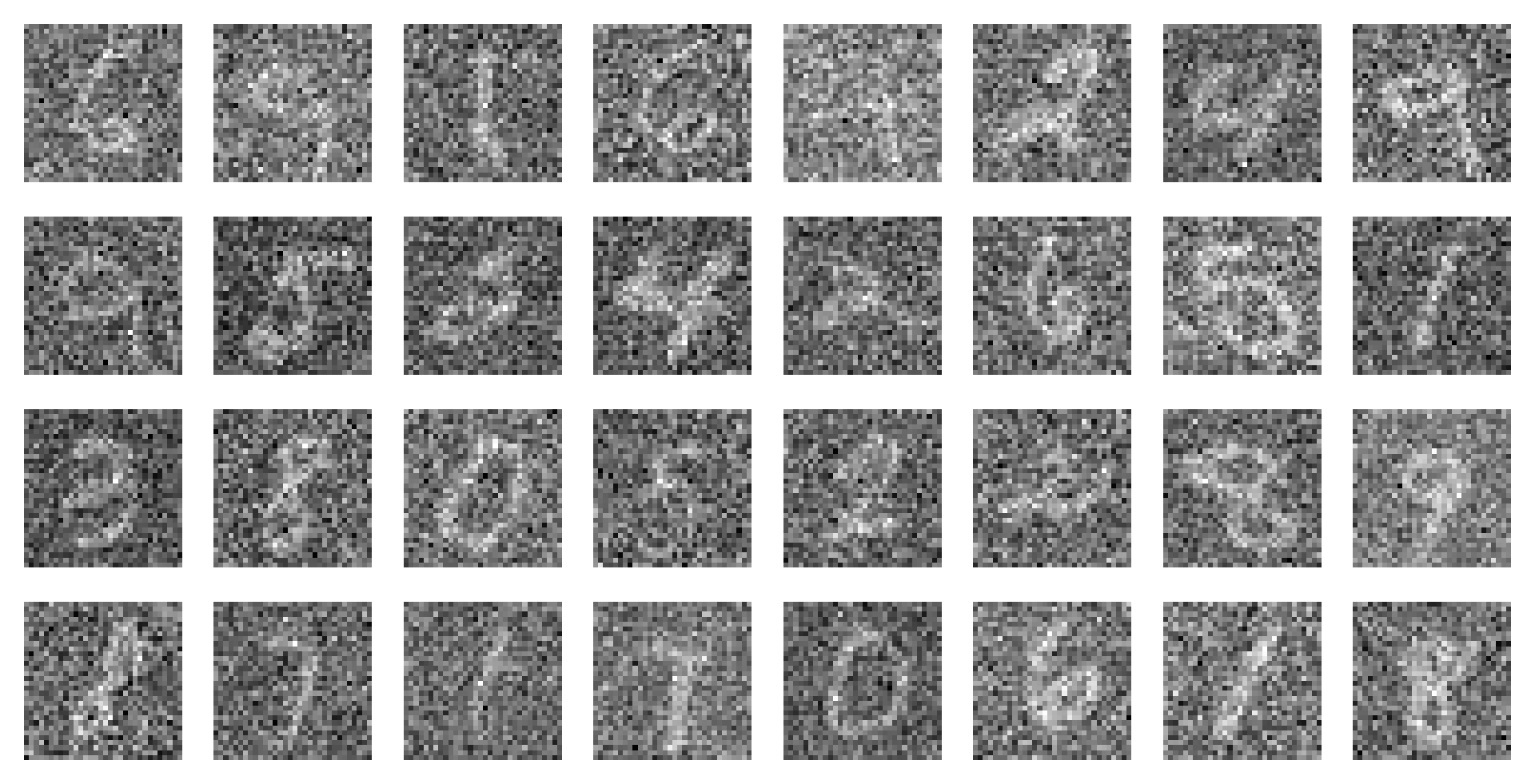}
        \caption{Degraded images $\widehat{\bm x}$}
    \end{subfigure}
    \begin{subfigure}{0.325\textwidth}
        \centering    
        \includegraphics[width=\linewidth, trim={0.0in 0.0in 0.0in  0.0in},clip]{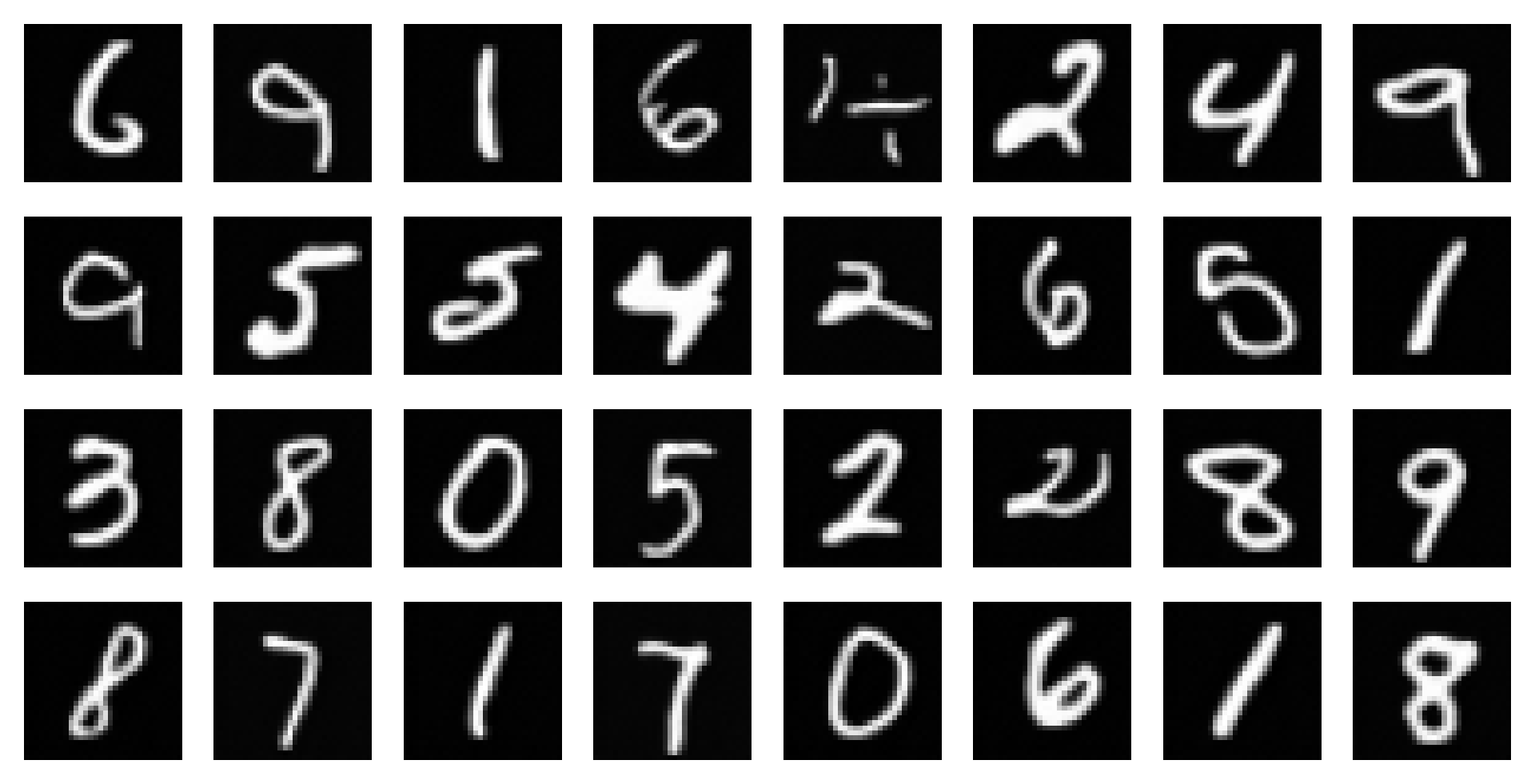}
        \caption{Conditional DDPM's output}
    \end{subfigure}
    \caption{ Data visualization for the performance evaluation  of the proposed scheme for image reconstruction enhancement. From left to right, the ground-truth information signals  $\boldsymbol{x}_0$, the degraded version $\widehat{\bm x}$ of the signals decoded at the output of the Sionna's simulator (before DDPM-based reconstruction), and the final reconstructed  signals via our method are  shown, respectively.   
    }
    \label{fig:vis_conv_ddpm}
    \vspace{0mm}
\end{figure*}

\begin{figure*}
\centering
    \begin{subfigure}{0.48\textwidth}
        \centering    
        \includegraphics[width=\linewidth, trim={0.0in 0.0in 0.0in  0.0in},clip]{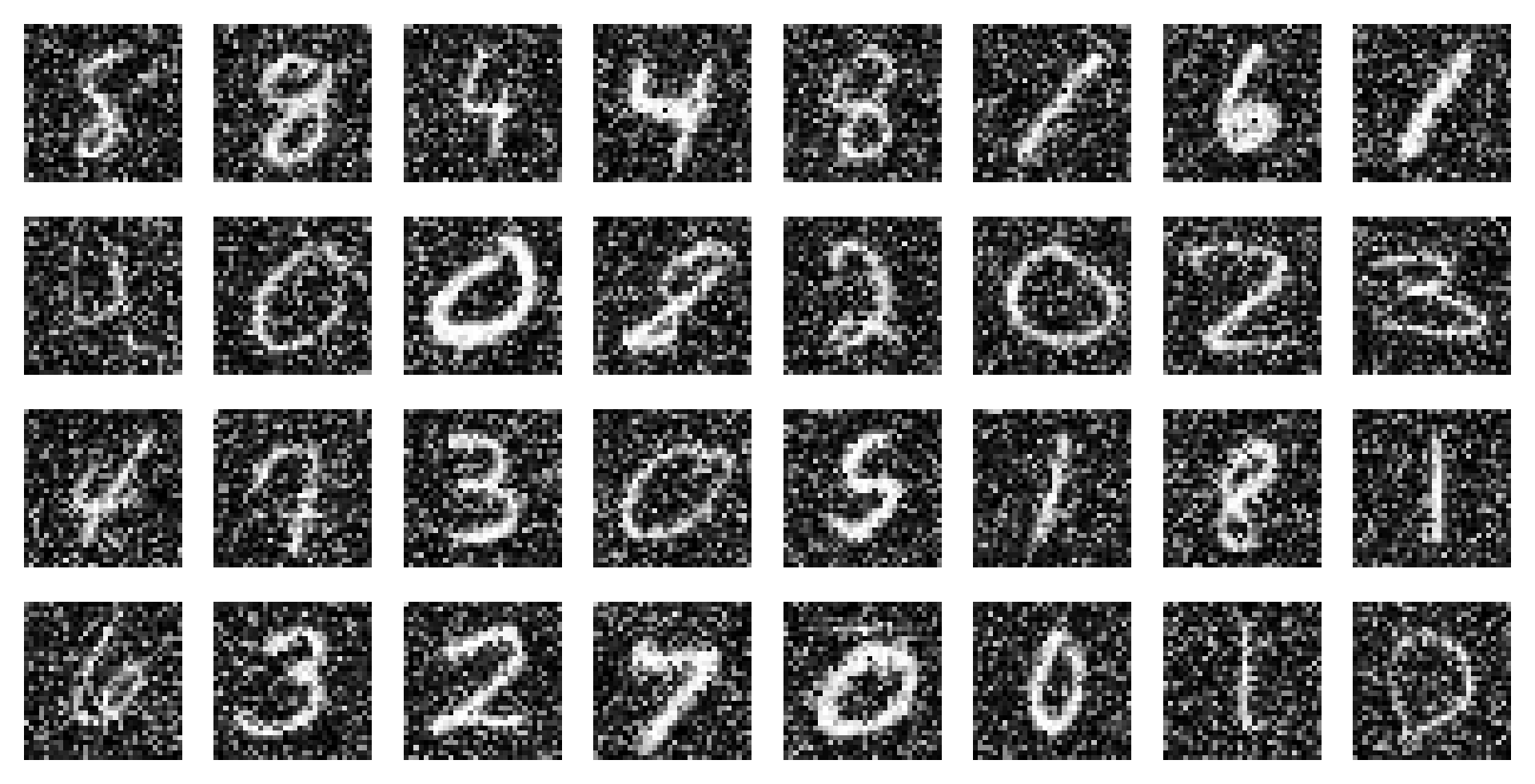}
        \caption{DNN-based receiver's output}
    \end{subfigure}
     \begin{subfigure}{0.48\textwidth}
        \centering    
        \includegraphics[width=\linewidth, trim={0.0in 0.0in 0.0in  0.0in},clip]{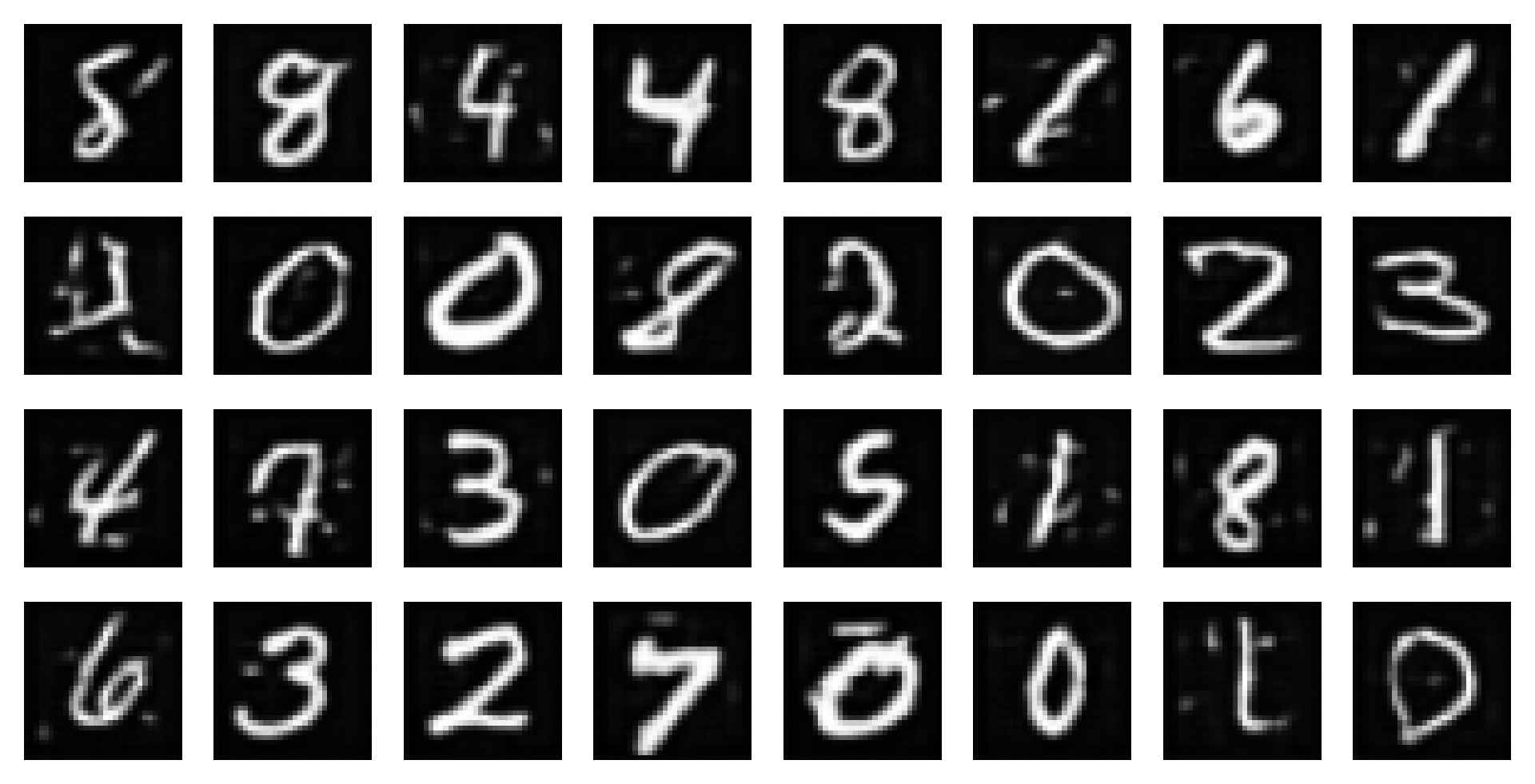}
        \caption{Conditional DDPM-aided receiver's output} 
    \end{subfigure}
    \caption{Comparison between the DNN-based receiver \cite{DNN} and our DDPM-aided model for image reconstruction.    
    } 
    \label{fig:vis_dnn_ddpm}
    \vspace{0mm}
\end{figure*}

\textcolor{black}{
\begin{itemize}
    \item DNN-based  receiver \cite{DNN} incorporated into Sionna's simulator\footnote{\url{https://nvlabs.github.io/sionna/examples/Sionna_tutorial_part2.html}}. This baseline is studied for both qualitative and  quantitative comparisons in Figs.  \ref{fig:vis_dnn_ddpm} and \ref{fig:psnr_DNN_DDPM}, respectively.   
For the DNN-based receiver baseline, we are inspired by the network  architecture proposed in \cite{DNN}, and   
since we have employed  our diffusion model at the receiver,   
we only exploit the receiver DNN of \cite{DNN} and fine-tune it for benchmarking, so that we can have a fair comparison. 
Three hidden linear layers with $64$ neurons and rectified linear unit (ReLU) activation functions are implemented, trained for  5000 iterations with learning rate $0.01$.  
{Based on our \textcolor{black}{trial  and errors},  increasing the depth of the conventional DNN, or the number of hidden neurons does not result in  any significant  improvement in its performance.} 
We show that our proposed model outperforms this baseline since the CDiff exploits 
the generative prior knowledge regarding  the underlying distribution of the ground-truth information signals. 
However, the DNN receiver does not
take the ``content'', nor any prior knowledge of the information
signals into account when decoding the information bits.  
\item VAE-based reconstruction.  
VAEs as another state-of-the-art 
GenAI-based decoding approach,  which are famous for image reconstruction, are compared with our DDPM approach.    
This is quantitatively studied in Fig.  \ref{fig:psnr_train}.
As mentioned earlier,  our approach  considers the ``contents''  of the transmitted signal during denoising and reconstruction.  
Therefore, to have a fair comparison from the perspective of ``prior knowledge'' utilization,   
we consider this baseline, which  also considers the  prior knowledge about  the underlying distribution of the ground-truth when reconstructing.       
The VAE's loss function is composed of two parts, i.e., the reconstruction loss and the Kullback-Leibler (KL) divergence loss. The former  measures the difference between the original data  and the reconstructed output  generated by the decoder, while the latter  measures the difference between the learned probability distribution and the predefined prior distribution.   
For the ``contracting path,'' four 2D convolutional layers  (\texttt{Conv2D}) are implemented with $32$, $64$, $64$, and $64$ filters, respectively, with kernel size of $3\times 3$, and ReLU activation functions,     
followed by a flattening layer  and a dense layer of size $32$ (to combine the down-sampled learned features before entering the latent space).  
Two dense layers further generate the parameters of the Gaussian distribution in the latent space (a.k.a. the mean vector $\mu$ and log variance vector $\sigma$ which represent   the compressed latent features of the input),   with latent space dimension $\texttt{Dim}$ set to $5$ and $20$ for different scenarios (see Fig. \ref{fig:psnr_train}).   
The decoder reconstructs the images  from the latent representation,  by exactly mirroring the encoder in reverse, by progressively up-scaling the compressed data back to its original dimension.   
Accordingly, the architecture is implemented  by a dense layer  and a reshape layer to up-scale to a higher dimension,  
and then followed by de-convolutional layers (\texttt{Conv2DTranspose}), to 
further upscale and refine  the data to reconstruct the original input.   We show the superior performance of our approach compared this baseline, thanks  to the high-quality sampling characteristic of diffusion models.  
\item Conventional Receiver (Naive scheme). This baseline is studied for qualitative comparison (data visualization in Fig. \ref{fig:vis_conv_ddpm}), as well as quantitative comparisons (Figs. \ref{fig:mse_coded} and \ref{fig:mse_coded_noCode}).    
Our proposed scheme  complements this baseline in a way that  instead of employing error correction codes with low 
code rates to improve data decoding  in extreme channel conditions, our proposed 
scheme can be employed while preserving the data rate.  
We show that we can maintain higher  data rates while also improving the reconstruction quality.   
We also show that
the functionality of the proposed CDiff 
can be exploited to compensate for the underlying errors 
within the noisy data at the receiver, outperforming the channel codes in conventional receivers. 
\end{itemize}
}
\textcolor{black}{
In what follows, we show how our proposed method outperforms or complements the 
existing approaches. 
}

\subsection{\textcolor{black}{Qualitative Comparisons:} Data Visualization}\label{ssec:evaluations_train} Fig. \ref{fig:vis_conv_ddpm} visualizes the performance of our conditional DDPM-aided reconstruction enhancement scheme. For this figure, we consider the extreme condition of $\Gamma = 0$ dB channel SNR, with the 
transmitter's  impairment level $\kappa^t = 0.1$ and the receiver's  impairment level of $\kappa^r = 0.15$. 
As can be seen from the figure, under such extreme conditions,   the naive format of a communication system cannot perform well in removing  the noise and distortions to obtain a clean denoised version of the ground-truth information signals 
(Fig. \ref{fig:vis_conv_ddpm}-(b)). 
In such scenarios, our proposed diffusion-aided scheme can improve the reconstruction of images by removing the noise and distortions conditioned on the degraded signals. This  results in a significant improvement in the decoding performance.  This improvement is actually obtained via exploiting the prior knowledge learnt by our conditional DDPM. 
One might argue that in such cases, we can play with the coding rate and employ  stronger error correction codes with lower code rates. However, this can significantly reduce the data rate, while stronger error correction codes might not  necessarily work well under such extreme conditions either. In the subsequent figures, we numerically study this and we show that the DDPM outperforms those schemes in low SNR regimes.   

Fig. \ref{fig:vis_dnn_ddpm} visualizes the output of our proposed diffusion-aided  scheme, compared to a DNN-based receiver as benchmark \cite{DNN}. 
For this experiment, we set  $\Gamma = 5$ dB SNR and the 
transmitter's  impairment level $\kappa^t = 0.1$ and the receiver's  impairment level of $\kappa^r = 0.15$.   
Notably,  compared to the DNN benchmark  our scheme highlights a considerably better performance in obtaining clean samples of the information signals, by removing noise and other residual decoding errors.  This is due to the fact that the diffusion-based method exploits the generative prior knowledge about the underlying distribution of the ground-truth information signals. However, a DNN-based receiver as proposed in \cite{DNN} does not take the ``content'', nor any prior knowledge of the information signals into account when decoding the information bits.       
We further emphasize that numerical comparisons between   our scheme  and the DNN benchmark  are provided  in the subsequent figures.


\subsection{\textcolor{black}{Quantitative Comparisons:} Reconstruction Performance}\label{ssec:evaluations_performance}
In this subsection, we numerically assess  the reconstruction  performance of the  proposed scheme in  different scenarios.

\begin{figure} 
\includegraphics
[width=3.45in, height=2.65in, trim={0.1in 0.1in 0.0in  0.0in},clip]{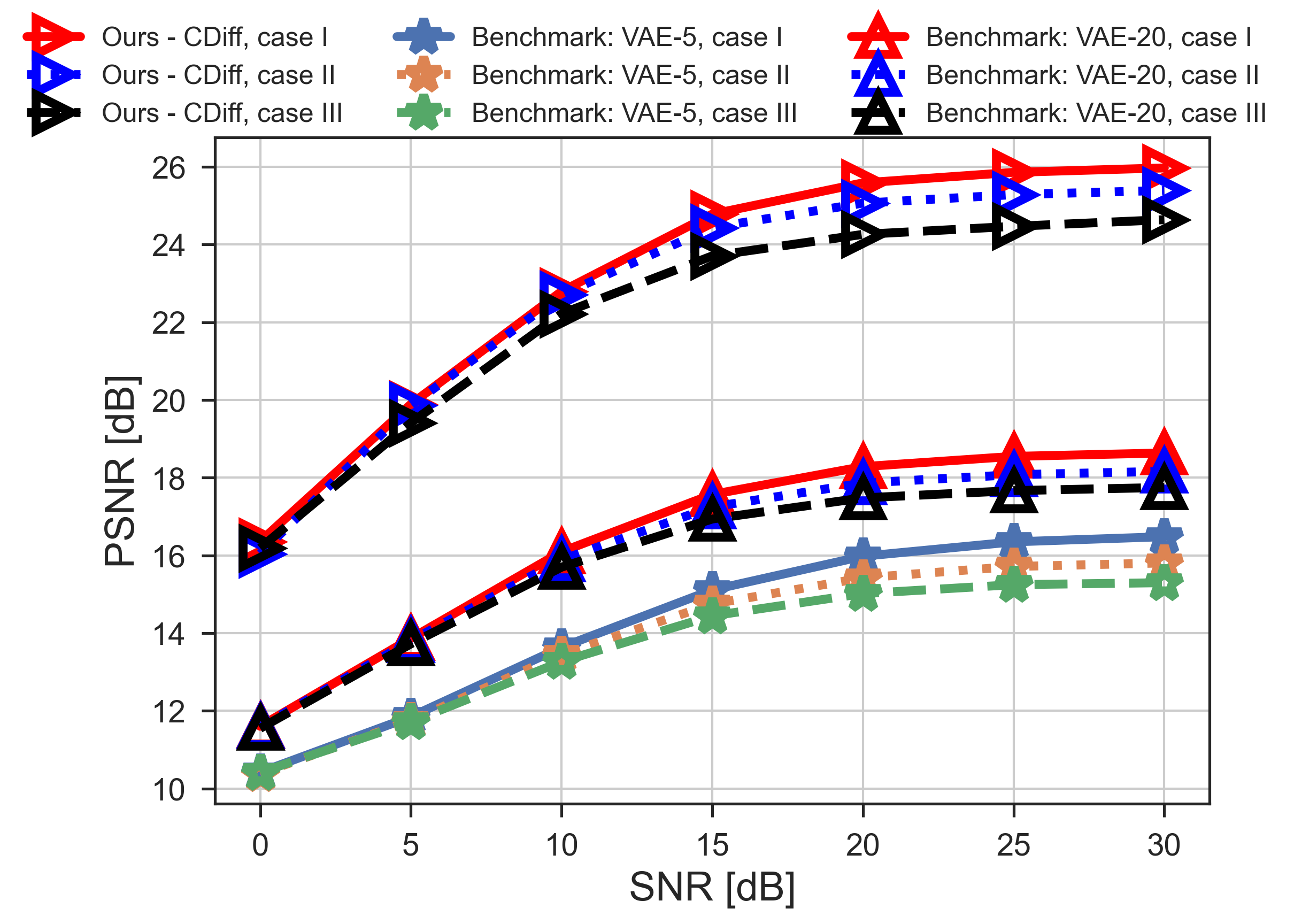}
\vspace{-3mm}
\caption{\textcolor{black}{Reconstruction performance in terms of PSNR for DDPM vs. VAE.}} 
	\label{fig:psnr_train} 
 \vspace{0mm}
\end{figure}

\begin{figure} 
\includegraphics
[width=3.65in, height=2.65in, trim={0.1in 0.0in 0.0in  0.0in},clip]{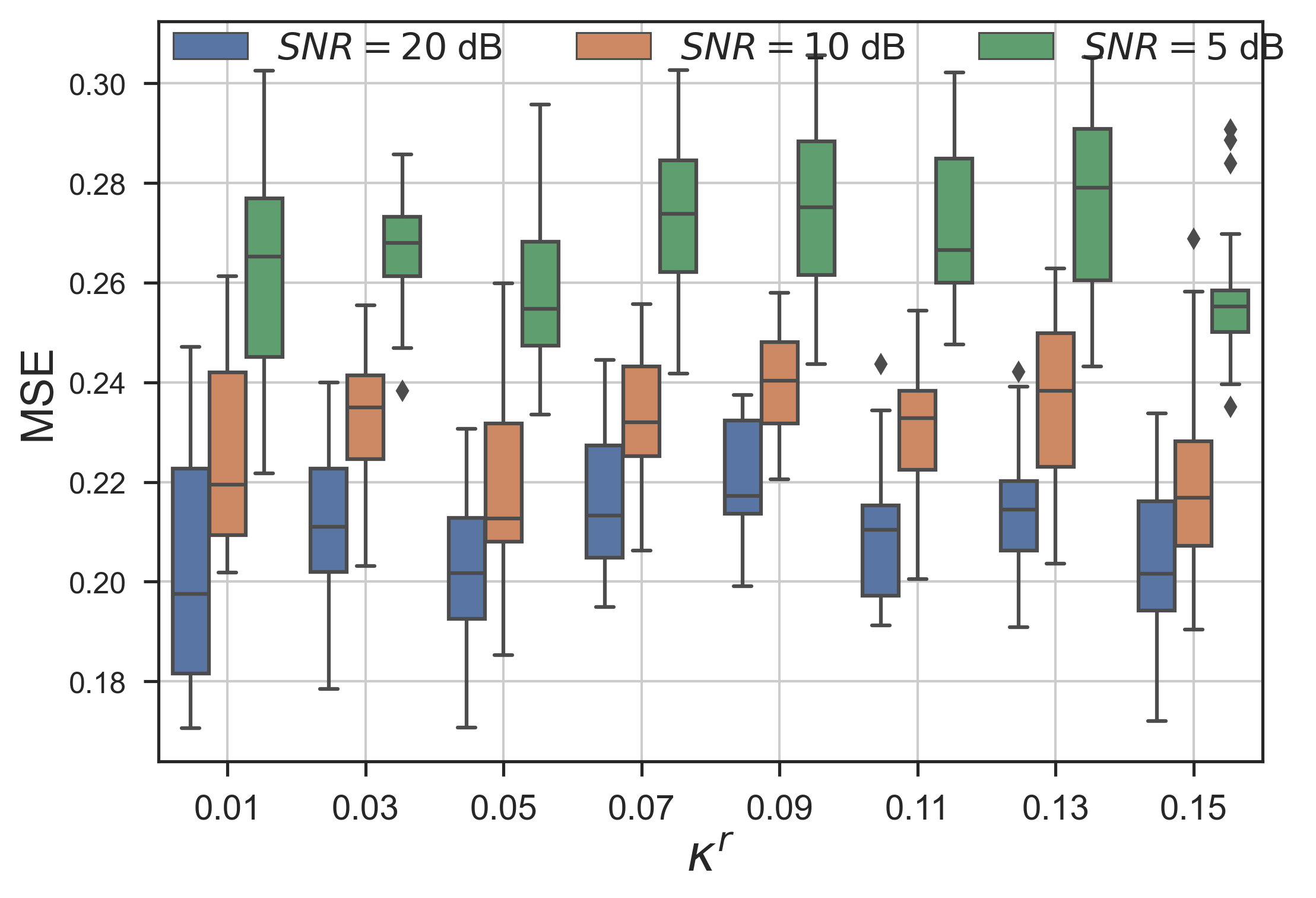}
\vspace{-5mm}
\caption{MSE between the original images  and the reconstructed ones   for different levels of hardware impairment.} 
	\label{fig:boxplot}
 \vspace{0mm}
\end{figure}

\textcolor{black}{Fig. \ref{fig:psnr_train} shows the reconstruction performance in terms of PSNR metric for our proposed  conditional DDPM-based approach, compared to the VAE-based reconstruction baseline which was introduced in Section \ref{subsec:baselines}.}  For case I to III, we consider $(\kappa^t, \kappa^r) = (0.01, 0.01)$, $(\kappa^t, \kappa^r) = (0.05, 0.1)$, and $(\kappa^t, \kappa^r) = (0.05, 0.15)$, respectively.  
\textcolor{black}{In the figure, VAE-$\texttt{Dim}$ shows the latent space dimension of the implemented VAE, where higher latent dimensions can offer a slightly better reconstruction quality.}   
\textcolor{black}{As can be seen from the figure, our proposed DDPM outperforms the VAE baseline with more than $40\%$ improvement in PSNR metric.  This is due to the high-quality sampling characteristic of diffusion models, while VAEs lack this important generative feature \cite{Trilemma}.   
To elaborate, in contrast to VAEs, diffusion models maintain high sample quality and strong mode coverage, 
because of  their unique feature for \emph{fine-grained denoising} which is realized via the reverse diffusion process.  This makes them suitable for the applications that interact with end-users and require high quality 
data reconstruction and generation.  
}
We further note that the saturation  observed in the PSNR trends over SNR is due to the so-called ``ceiling phenomena'' in the  literature of  hardware-impaired communications \cite{HI}. To elaborate,  by increasing the SNR, the effect of channel noise becomes negligible compared to the effect of residual impairments within the RF chain of communication systems, hence the performance metrics become saturated \cite{HI}.    
As can be seen from the figure, the PSNR increases with the  increase in SNR, due to having less noise communication, which makes it easier  for the employed diffusion to denoise and reconstruct when facing with a less degraded signal. In addition, lower impairment levels indicate higher  qualities for the transceiver hardware which results in  less mismatches  between the transmitted signals and the decoded ones, thus  improving the reconstruction   performance. Generally speaking,  the proposed scheme shows a \emph{near-invariant}  performance with respect to the hardware impairments (as we can see the slight change in PSNR with respect to the three scenarios). This is further studied in Fig. \ref{fig:boxplot}.

\begin{figure} 
\includegraphics
[width=3.4in, height=2.75in, trim={0.1in 0.0in 0.0in  0.0in},clip]{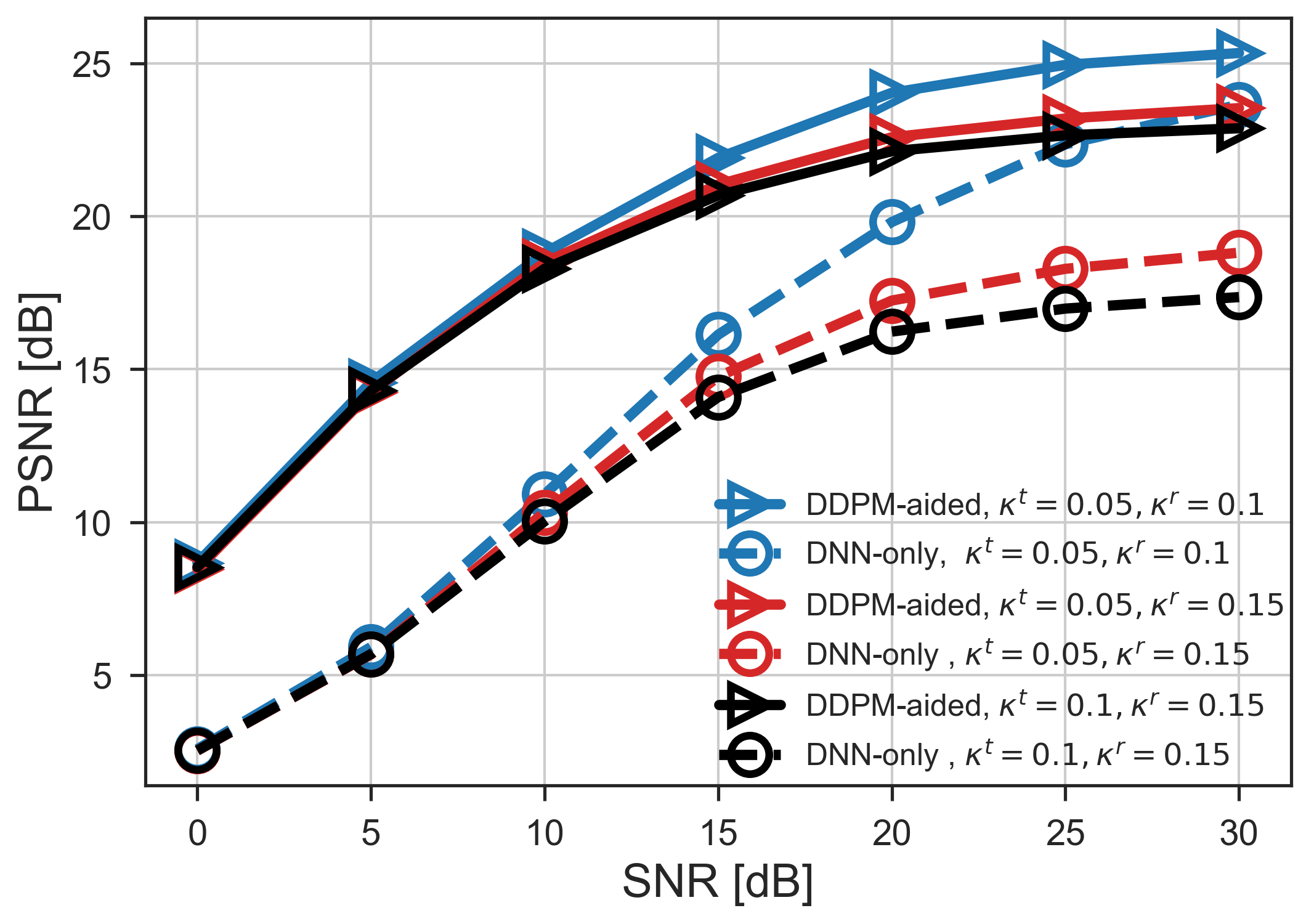}
\vspace{0mm}
\caption{Reconstruction performance in terms of PSNR: Comparison between generative (DDPM-based) learning approach vs. discriminative   (DNN-based) learning approach.} 
	\label{fig:psnr_DNN_DDPM} 
 \vspace{0mm}
\end{figure}

 
\begin{figure} 
\includegraphics
[width=3.4in, height=2.5in, trim={0.1in 0.0in 0.0in  0.0in},clip]{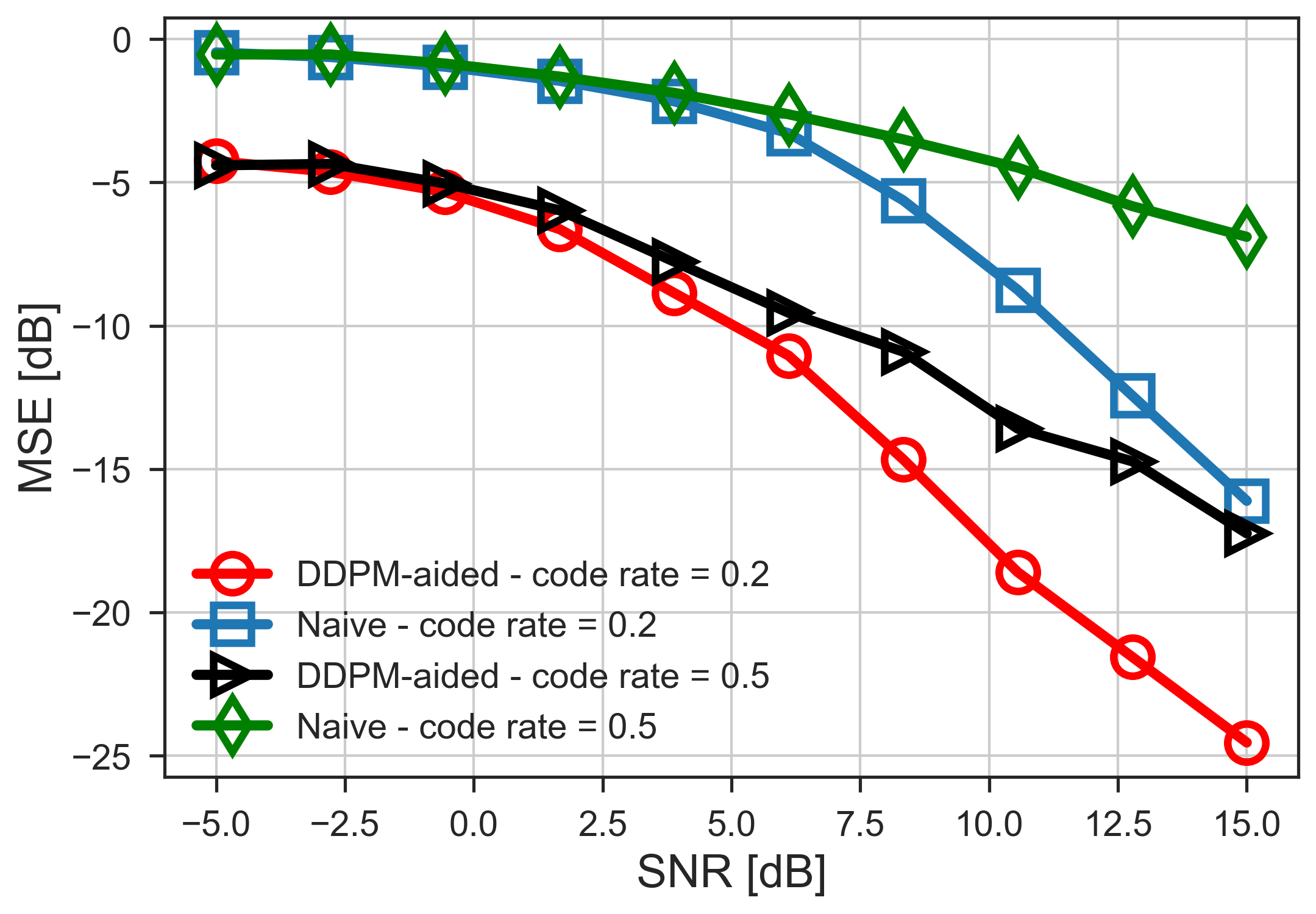}
\vspace{0mm}
\caption{Reconstruction performance in terms of MSE between the original images and the reconstructed ones.}
	\label{fig:mse_coded}
 \vspace{0mm}
\end{figure}

Fig. \ref{fig:psnr_DNN_DDPM} evaluates the reconstruction performance of our proposed  approach when it is employed in conjunction with the Sionna's neural receiver\footnote{\url{https://nvlabs.github.io/sionna/examples/Sionna_tutorial_part4.html}}, compared to the non-DDPM approach which simply employs the Sionna's neural receiver without any diffusion enhancement.      
The figure highlights about $10$ dB improvement  in low-SNR regimes, as well as more than $5$ dB improvement in higher SNRs,  when  using our generative model. This is due to the fact that our  DDPM-based scheme takes into account the entire underlying distribution of the ground-truth signals that are supposed to be sent over the wireless channel. However, the conventional DNNs  neglect the underlying ``content'' of the information signals, and simply care about the ``bit-wise'' transmission, using discriminative loss  functions, such as binary cross-entropy,  rather than a content-wise'' approach.

Fig. \ref{fig:mse_coded} demonstrates the image  reconstruction error, in terms of the mean squared error  (MSE)  versus SNR for $\kappa^t = \kappa^r = 0.25$.  The figure shows that under extreme conditions  (low-SNR and mid-SNR regimes and highly-impaired communications), a naive scheme (without our proposed diffusion model) cannot perform well in terms of the image reconstruction, even if the error correction rate is decreased from $0.5$ to $0.2$.  When the conditional DDPM is employed, the reconstruction is enhanced significantly, by more than 10 dB. This is due to the fact that our scheme  leverages the  denoising  capabilities of diffusion models, as well as being conditioned on the degraded output of a naive communication system, which guides the diffusion to improve the data decoding and reconstruction as shown in the figure.   
The figure suggests that instead of employing error correction codes with lower code rates, to improve the data decoding quality,  our proposed scheme can be employed while preserving the data rate. As can be seen from the figure, more than $5$ dB performance improvement, as well as $2.5$ times higher data rates can be achieved under low SNR regimes with our proposed scheme, instead of reducing the code rate from $0.5$ to $0.2$ in a naive scheme. This is applicable to scenarios in which data rates and reliable communications are assumed to be taken into account at the same time. 
Also note that although the gap between a naive scheme with lower code rate (blue curve) and the DDPM-aided scheme with code rate of $0.5$ (black curve) decreases as the SNR increases, the figure shows that if we employ the DDPM, the same performance can be achieved, while we do not even need to decrease the data rate.

\begin{figure} 
\includegraphics
[width=3.4in, height=2.55in, trim={0.1in 0.0in 0.0in  0.0in},clip]{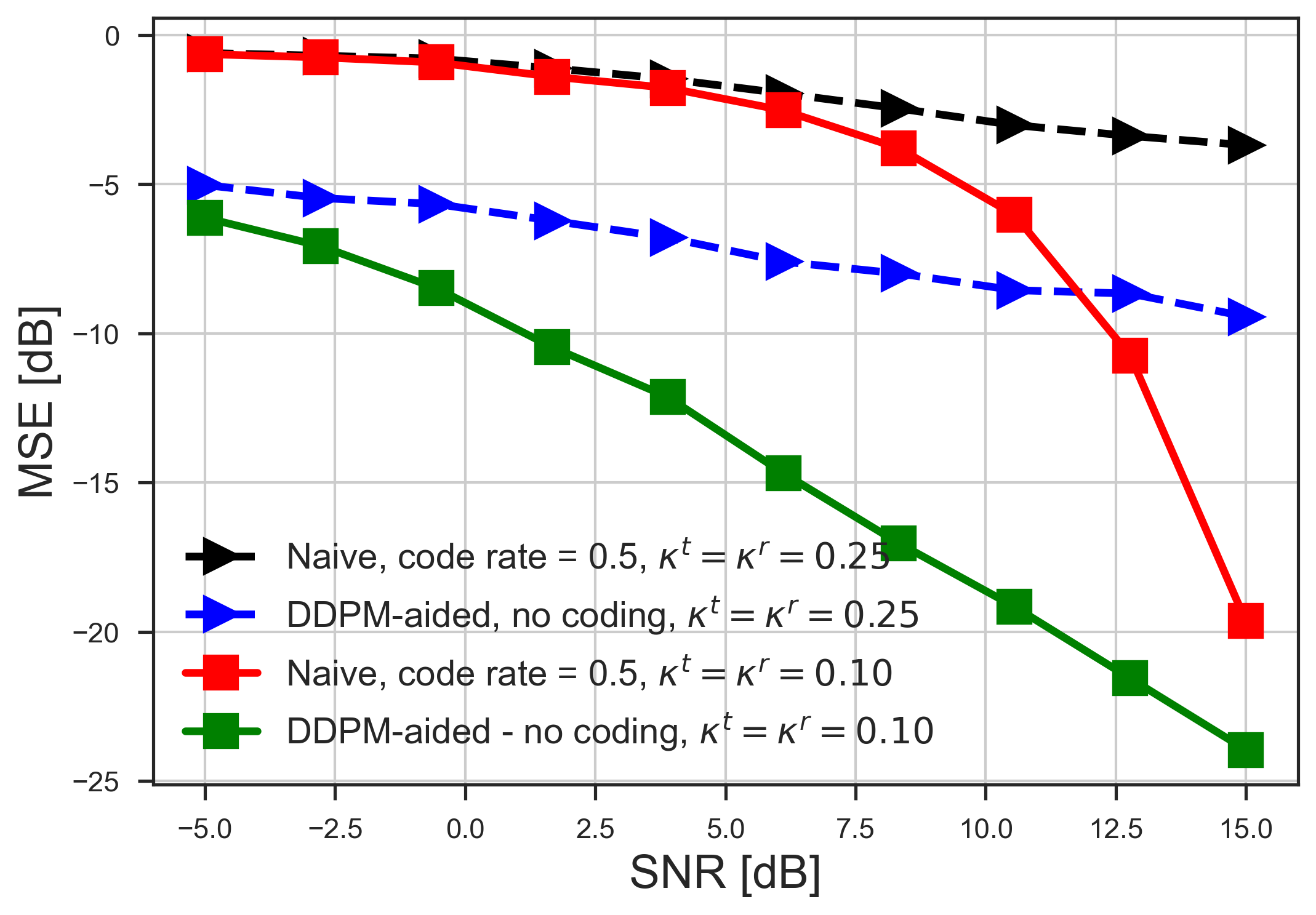}
\vspace{0mm}
\caption{Reconstruction performance between the original images and the reconstructed ones for different impairment levels: Diffusion models vs. channel codes.}
	\label{fig:mse_coded_noCode} 
 \vspace{-4mm}
\end{figure}

Fig. \ref{fig:mse_coded_noCode} shows the reconstruction performance between the original
images and the reconstructed ones for two different impairment
levels. In this experiment, we aim to show that the functionality of the proposed CDiff  can be exploited to compensate for the underlying errors within the noisy reconstructed images obtained at the receiver, instead of employing channel codes. As can be seen from the figure,  by employing  our conditional DDPM, the quality of the information signals (images in our scenario) can be improved more than the case where error correction codes are employed. 
For instance, by employing our proposed scheme,  more than $10$ dB improvement in reconstruction  performance can be achieved at $5$ dB SNR and impairment level $0.1$.  This can be explained due to the fact that channel codes do not take into account the underlying structure and the contents of the information  signals that are transmitted over the channel. However,  our proposed scheme essentially considers the content of the transmitted  signals    when carrying out the denoising and reconstruction.  Also note that error correction codes require sending redundancy bits which decrease the data rate, while our diffusion-based scheme has achieved better performance without the need for transmitting such redundancy bits. 
Therefore, the figure indicates that our scheme can improve the reliability of the communication systems, while saving the resources for only transmitting information bits.   
\textcolor{black}{Obviously, this performance improvement comes with a trade-off in the sense that the DDPM model requires higher computation resources than the error correction codes. Nevertheless, one should note that the \emph{computations related to error correction codes are implemented in both of the communication sides, i.e., the transmitter (the encoder part), and the receiver (the decoder part), while the proposed diffusion is only employed and run at the receiver side.}}

\textcolor{black}{
\begin{remark}
\it
We remark that the outperformance of our scheme compared to the traditional channel codes comes with a trade-off.  
DDPMs may fall short  in sampling speed due to the stepwise denoising process.  Although this appears to limit the wide adoption of diffusion models for some applications, there already exists an active line  of research  focusing  on accelerating the generation 
process.  Such computation-efficient  variants of diffusion framework, such as  \cite{wcl}, can help  facilitate the applicability  of diffusion models  to a wide range of real-world wireless scenarios. 
\end{remark}}

\textcolor{black}{\subsection{\textbf{Complexity Analysis}}
We further address the potential computational overhead of implementing DDPMs to highlight the complexity-performance  trade-off compared with the  traditional error correction coding  methods.   
\subsubsection{Traditional channel codes} 
According to the coding theory, 
the encoding of LDPC codes is typically performed using a generator matrix 
$\mathbf{G}$ that transforms the input information bits into codewords via multiplying the input stream by the 5G LDPC matrices.  The complexity of encoding an LDPC code is proportional to the number of non-zero entries in the generator matrix, and is approximated by $\mathcal{O}(n)$, where $n$ is the length of the codeword. 
The complexity of LDPC decoding is more computation-intensive than encoding,   because of the \emph{iterative} algorithms of decoding. 
According to coding theory \cite{coding}, the complexity of LDPC decoding  depends on the number of non-zero entries in the parity-check matrix $\mathbf{H}$, which represents the edges in the bipartite graph used for message passing in the belief propagation (BP) algorithm \cite{coding, BP}. 
The BP algorithm iteratively updates the likelihoods (beliefs) for each bit, passing messages between the variable nodes (representing bits) and check nodes (representing parity checks) in a bipartite  graph.
Then the complexity per iteration is proportional to the number of edges, $E$,  in the  graph. 
For an LDPC with the 
length of the codeword denoted by $n$,  the complexity per iteration is approximately  
$\mathcal{O}(E)$, and since the number of edges $E$ is typically proportional to 
$n$ (due to the sparsity of $\mathbf{H}$), the computation complexity is $\mathcal{O}(n \cdot I)$, where  
the number of  iterations $I$ for LDPC decoding typically depends on the target error-correction performance. Assuming $F$ floating-point operations (FLOPs) per message update on an edge (for calculating likelihoods and probabilities), the total FLOPs for the traditional error correction  would be  
\begin{equation}
 \text{FLOPs}^\mathsf{LDPC} = \mathcal{O}(I \cdot n \cdot F). 
\end{equation}
\subsubsection{DDPM's complexity}
The computational complexity of a U-Net can generally be analyzed in terms of the number of FLOPs for its convolutional layers within the contractive and expansive paths  \cite{unet, cnn_complexity_calc}. 
Key parameters that influence the complexity are 
input size (considering the spatial dimensions of the input image  as 
$H\times W$ for height and width), and
depth of the U-Net architecture, $D$. This reflects  the number of down-sampling and up-sampling stages in the U-Net. 
For a U-Net with  depth  
$D$, the input size is halved at each down-sampling layer in the contracting path, and the number of channels doubles \cite{unet}.  
Hence, assuming $k\times k$ filter size and $C$ as the number of input channels, the FLOPs at each layer $i \in [D]$ of the contractive (down-sampling) path can be calculated as   
\begin{equation}
    \text{FLOPs}^{\mathsf{UNet, ds}}_i = k^2  \left( \frac{H}{2^i} \cdot  \frac{W}{2^i} \right) \cdot (2^{i-1} C) \cdot (2^i C).
\end{equation}
The reverse  happens in the up-sampling (expansive path), and hence we have 
\begin{equation}
     \text{FLOPs}^{\mathsf{UNet, us}}_i = k^2   \left( \frac{H}{2^{D-i}} \cdot \frac{W}{2^{D-i}} \right) \cdot (2^{D-i} C) \cdot (2^{D-i-1} C).
\end{equation}
In addition to up/down-sampling,  U-Net architecture also maintains the  skip connections, the complexity of which  
is 
negligible with respect to  that of the convolution layers. 
Hence, 
summing over all the down-sampling and up-sampling layers for a depth  
$D$ U-Net, the total complexity of the U-Net architecture can be approximated by \cite{cnn_complexity_calc}
\begin{equation}
\text{FLOPs}^{\mathsf{U-Net}} \approx \mathcal{O}(H\cdot W\cdot C^2 \cdot D). 
\end{equation} 
Recall that in the denoising and reconstruction process of our proposed CDiff method,  the U-Net is applied at each of the $T$ diffusion steps to gradually denoise the data. 
At each step, the complexity is that of running the U-Net once.
Therefore, the total complexity of the diffusion model is the complexity of the U-Net multiplied by the number of diffusion steps 
\begin{equation}
\text{FLOPs}^{\mathsf{DDPM}} \approx \mathcal{O}(T \cdot H\cdot W\cdot C^2 \cdot D). \label{eq:DDPM_complex}
\end{equation}
This indicates that the complexity grows linearly with the depth 
$D$ of the network, and linearly with the spatial size of the input $H\times W \times C^2$. 
}  
\textcolor{black}{
\subsubsection{Comparison} 
If we assume the number of denoising  steps $T$ to be equivalent/counterpart to the number of decoding iterations, $I$, in traditional decoding,  and the input spatial dimensions $H\times W \times C^2$ in the  DDPM method as the equivalent counterpart to the size of the input data-stream to the LDPC error correction decoding (input codeword length),   
the computation complexity of the DDPM-based approach depends on how deep the neural  architecture of the diffusion framework is.  
This is reflected as the  depth $D$ of the network in \eqref{eq:DDPM_complex}. 
This shows that, similar to any other AI/ML  models applied to wireless problems,  the complexity overhead of diffusion-based  models is also dominated  by nothing except the underlying  neural network architecture.  
This also highlights the importance of studying efficient implementations of diffusion architectures and  quantization and pruning techniques for the wide adoption of such models in real-world wireless problems.  
}

Finally, Fig. \ref{fig:boxplot} studies the effect of different impairment  levels on the reconstruction performance of our scheme over Rayleigh fading channels. For this figure, we set $\kappa^t = 0.05$ and vary the impairment level  of the receiver, $\kappa^r$,  over the typical ranges specified in Ref. \cite{HI}. The reconstruction results are obtained in terms of MSE metric over $20$ realizations of the system.  The figure  highlights an important characteristic  of our proposed scheme. Our DDPM-based communication system  is \emph{robust} against hardware distortions, since  the reconstruction performance does not change with the increase in the impairment  level.  
Notably, our scheme
showcases a \emph{near-invariant} reconstruction performance with
respect to noise and impairment levels. 
This  is achieved  due to the  so-called ``variance scheduling''  of diffusion framework in \eqref{eq:fwd_sample_gen_diffusion}, which  allows the system to become robust against a wide range of distortions caused by channel and hardware impairments.      

\subsection{\textcolor{black}{Further analysis and discussions on model's  robustness}} 
\textcolor{black}{We would like to provide some further discussions and highlight how the proposed CDiff method offers a robust  performance  under varying conditions,  demonstrating  its reliability in practical scenarios.
The fact that the model adapts to varying noise conditions and is effective  in low-SNR regimes can be addressed  by the  compilation of the following reasons.} 

\textcolor{black}{
\begin{itemize}   
\item DDPMs maintain the \textbf{variance scheduling} in the forward diffusion mechanism, wherein the samples are diffused at each time-step {$t \in [T]$} 
according to $q(\boldsymbol{x}_t \vert \boldsymbol{x}_{t-1})$ in (1).    
    Invoking  (3)--(6), one can  model  the  forward
diffusion according to a ``diffusion SNR''  
$    \gamma^{\mathsf{DDPM}}(t) = \frac{1-\bar{\alpha}_t}{\bar{\alpha}_t}.$
The mathematical relations between the diffusion SNR, and the communication SNR would be an interesting direction to study  in  the future.    
However, the important point here is that in practice, the receiver of a communication system might not have any prior knowledge regarding  the levels of RF impairments and communication SNRs. The  forward diffusion  does not consider any form of pre-defined hardware impairment levels   
or communication  SNRs either.  
This is exactly the point of forward diffusion in our scheme. We purposefully  ``diffuse'' Gaussian noise which maintains the maximum entropy, thus maximum uncertainty from the information-theoretic perspective, and then  the DDPM tries to make itself ``robust'' against a wide range of  uncertainties.      
This intuitively reflects a robust performance  over a wide range of scenarios  
as shown in the paper.       
\item  The concept of  \textbf{conditioning} in our  scheme offers an additional level of robustness for the system.    
   We incorporated the  noisy decoded data  $\widehat{\bm{x}}$ into the diffusion framework to learn to remove the exact noise from $\widehat{\bm{x}}$ and recover the ground-truth $\bm{x}$.  This way, the CDiff learns to estimate both the Gaussian noise $ \epsilon$ and the residual errors ${\widehat{ x}}-{ x}_0$. 
Moreover, this  further allowed  us to consider different communication  SNRs during the training as described  in \ref{subsec:Setup}.  
Hence, this conditioning has helped us incorporate the conditions  of a practical wireless
system into account, 
making  the system more robust to different  conditions in the inference phase.  
\item 
The notion of \textbf{time embedding}  in the DDPM also reinforces the model robustness.  
Considering the variance scheduling $\beta_t$,   
each time-instance has a unique mapping to a specific noise variance. Hence, intuitively speaking, each  time-step equivalently corresponds to a specific level of
 noise-plus-distortion in the batch of received signals. 
Incorporating the time-steps {$t \in [T]$} into our neural network as explained  in Section  \ref{subsec:Setup}, makes the model ``know''  at which particular time-step  is operating for each level of channel noise, making it robust against different SNR and/or impairment  levels.    
This notion $t$ is also reflected in the loss function of our model in (35),     
so  data 
samples with different levels of ``noisiness'' are processed  at different time-steps during  training. 
\end{itemize}
}

\section{Conclusions}\label{sec:concl} 
In this paper, we have proposed conditional DDPMs  to  enhance the data reconstruction in wireless
communication schemes. The key idea is to leverage
the generative prior of diffusion models to  learn a noisy-to-clean transformation of the information signal, conditioned on the degraded version of the receiver data. 
The proposed scheme is applicable to scenarios in which a prior knowledge of the 
information signals is available, e.g., in multimedia transmission.
Our numerical results on MNIST dataset have shown that instead of employing complicated channel codes that 
reduce the information rate, CDiffs  can be employed that improve 
reliable data transmission, especially under extreme channel
conditions due to low SNR, or hardware-impaired communications. 
Our evaluations have been carried out using 
the NVIDIA Sionna simulator, where it has been shown that more than $10$ dB improvement  as well as $2$ times higher data rates can be
achieved under low SNR regimes with our proposed scheme, compared to a conventional error correction-based data transmission and reconstruction.  Moreover, we highlight the outperformance of the proposed generative learning approach compared to the conventional contrastive learning models based on DNNs.  


\end{document}